\documentclass[10pt, conference, onecolumn]{IEEEtran}

\usepackage{mdwmath}
\usepackage{amsmath}
\usepackage{amssymb}
\usepackage{cite}
\usepackage{epsfig}
\usepackage{url}

\newtheorem{rem}{Remark}
\newtheorem{theo}{Theorem}

\newtheorem{lem}{Lemma}

\newcommand{\rank}{\text{rank}}

\newcommand{\diag}{\text{diag}}

\newcommand{\Span}{\text{span}}

\newcommand{\eqdef}{\stackrel{\triangle}{=}}

\newenvironment{example}[1][Example]{\begin{trivlist}
\item[\hskip \labelsep {\bfseries #1}]}{\end{trivlist}}

\begin{document}
\author{Dimitris S. Papailiopoulos and Alexandros G. Dimakis\\
Electrical Engineering\\
University of Southern California\\
Los Angeles, CA 90089\\
Email:\texttt{\{papailio, dimakis\}@usc.edu}
\and
Viveck R. Cadambe\\
Electrical Engineering and Computer Science\\ 
University of California Irvine, \\
Irvine, CA, 92697\\
Email: \texttt{vcadambe@uci.edu}
}

\title{Repair Optimal Erasure Codes through Hadamard Designs}
\maketitle

\begin{abstract}
In distributed storage systems that employ erasure coding, the issue of minimizing the total {\it communication} required to exactly rebuild a storage node after a failure arises.
This repair bandwidth depends on the structure of the storage code and the repair strategies used to restore the lost data.
Designing high-rate maximum-distance separable (MDS) codes that achieve the optimum repair communication has been a well-known open problem. 
In this work, we use Hadamard matrices to construct the first explicit $2$-parity MDS storage code with optimal repair properties for all single node failures, including the parities. Our construction relies on a novel method of achieving perfect interference alignment over finite fields with a finite file size, or number of extensions. 
We generalize this construction to design $m$-parity MDS codes that achieve the optimum repair communication for single systematic node failures and show that there is an interesting connection between our $m$-parity codes and the systematic-repair optimal permutation-matrix based codes of Tamo {\it et al.} \cite{Tamo} and Cadambe {\it et al.} \cite{PermCodes_ISIT, PermCodes}.
\end{abstract}

\section{Introduction}
Distributed storage systems have reached such a massive scale that recovery from failures is
now part of regular operation rather than a rare exception \cite{GFS}.
Large scale deployments typically need to tolerate multiple failures, both for high availability 
and to prevent data loss. Erasure coded storage achieves high failure tolerance without requiring 
a large number of replicas that increase the storage cost. Three application contexts where 
erasure coding techniques are being currently deployed or under investigation are Cloud storage systems~\cite{KhanBPH},
archival storage, and peer-to-peer storage systems like Cleversafe and Wuala.

One central problem in erasure coded distributed storage systems is that of maintaining an encoded representation when failuires occur.
To maintain the same redundancy when a storage node leaves the system, a {\it newcomer} node has to join the array, access some existing nodes, and exactly reproduce the contents of the departed node.
In its most general form this problem is known as the {\it Exact Code Repair Problem} \cite{DimakisGWWR:08}, \cite{storagewiki}. 
There are several metrics that can be optimized during repair: 
the total information read from existing disks during repair~\cite{ACTEMT,RDP_repair}, the total information communicated in the network (repair bandwidth~\cite{DimakisGWWR:08}), or the total number of disks required for each repair~\cite{Oggier11,KhanBPH, Tamo, PermCodes}. 

Currently, the most well-understood metric is that of repair bandwidth.
For designing $(n,k)$ MDS erasure codes that have $n$ storage nodes and can tolerate any $n-k$ failures, the information theoretic cut-set bounds for repair communication were specified in~\cite{DimakisGWWR:08} and shown to be achievable for all values of $n,k$ in a series of recent papers~\cite{ShahRKR,KVSKR:09,CadambeWinc,SuhCodes,RashmiProduct,Survey}. In particular, it was shown that for a $(n,k)$ code, if a single node fails, downloading $\frac{1}{n-k}$ fraction of every surviving disk is sufficient and optimal in terms of repair bandwidth for the repair of a failed node. Beyond MDS codes,\cite{DimakisGWWR:08} demonstrated a tradeoff between storage and repair communication, and code constructions for other points of this tradeoff are under active investigation, see e.g.~\cite{RashmiProduct,Survey}, or \cite{Shum} for multiple node repair schemes. On this tradeoff, the minimum storage point is achieved by MDS erasure codes with optimal repair, also known as Minimum Storage Regenerating (MSR) codes.

For code rates $k/n\leq 1/2$, explicit MSR codes were designed by Shah {\it et al.}~\cite{ShahRKR}, Rashmi {\it et al.}~\cite{RashmiProduct}, and Suh {\it et al.}~\cite{SuhR:09}. 
For the high-rate regime, however, the only known complete constructions~\cite{CadambeWinc,SuhCodes} require large file sizes (symbol extensions) and field sizes. 
These constructions use the symbol extension interference alignment (IA) technique of \cite{CadambeJ:08} to establish that there exist MDS storage codes, that come arbitrarily close to (but do not exactly match) the information theoretic lower bound for the repair bandwidth for all $n$, $k$. 
These asymptotic constructions are impractical due to the arbitrarily large finite field size and the fast growing file size, required even for small values of $n$ and $k$.

{\bf Our Contribution}: 
We introduce the first explicit high-rate $(k+2,k)$ MDS storage code with optimal repair communication.
Our storage code exploits fundamental properties of Hadamard designs and perfect IA instances that can be understood through the use of a lattice representation of the symbol extension technique of Cadambe {\it et al.} \cite{CadambeJ:08,CadambeWinc,SuhCodes}. Our coding and repair strategy bears resemblance to the notion of ergodic interference alignment \cite{Nazer_Ergodic}, which is a finite-symbol-extension based IA scheme in the wireless channels.

Independently of this work, there has recently been a substantial progress in designing high-rate explicit MSR codes. Tamo et al.~\cite{Tamo} and Cadambe et al.~\cite{PermCodes} designed 
MDS codes for any $(n,k)$ parameters that have optimal repair for the systematic nodes, but not the code parities. 
It seems that extending these designs to allow optimal parity repair is not straightforward.
The advantage of our work is that all $n$ nodes are optimally repaired and the disadvantage is that our construction is currently only optimal for $n-k=2$. 

Our key technical contribution is a scheme that achieves {\it perfect} interference alignment with a finite number of extensions. 
This was developed in~\cite{PD1} and used in $2$ parity storage code with optimal repair for $k$ nodes and near optimal repair of $2$ nodes, that can handle any single node failure. 
We use a combinatorial view of different interference alignment schemes using a framework we call dots-on-a-lattice. 
Hadamard matrices are shown to be crucial in achieving finite perfect alignment and ensuring the full-rank of desired subspaces. 

Finally, we present $m$-parity MDS code constructions based on Hadamard designs that achieve optimal repair for systematic node failures, but suboptimal repair for parity nodes.
We show that these codes are equivalent to codes that involve permutation matrices in the manner of \cite{Tamo} and \cite{PermCodes} under a similarity transformation.

{\it $2$-parity Code Parameters:}
Assuming that the file to be encoded has size $M=k2^{k+1}$, each of the $k+2$ storage nodes stores a coded block of size $\frac{M}{k}$.
Repairing a single node failure costs $\frac{k+1}{2k}M$ in repair communication bandwidth, matching the theoretic lower bound. 
Finally, we give explicit conditions on the MDS property of the code and show that finite fields of size greater than or equal to $2k+3$ suffice to satisfy them.

{\it $m$-parity Code Parameters:}
For file sizes $M=km^k$, our $(k+m,k)$ codes achieve a repair communication bandwidth of $\frac{k+m-1}{mk}M$ for single systematic node failures, matching the information theoretic lower bound. 
The MDS property of these codes is shown to hold for arbitrarily large finite fields with high probability.

\section{MDS Storage Codes with $2$ Parity Nodes}
In this section, we consider the code repair problem for MDS storage codes with $2$ parity nodes. 
After we lay down the model for repair, we continue with introducing our code construction.
Let a file of size $M=kN$ denoted by the vector ${\bf f}\in\mathbb{F}_q^{kN }$ be partitioned in $k$ parts ${\bf f}=\left[{\bf f}^T_1\ldots{\bf f}^T_k\right]^T$, each of size $N$, where $N$ denotes the subpacketization factor, $\frac{N}{2}\in\mathbb{N}^*$.\footnote{$\mathbb{F}_q$ denotes the finite field, over which all operation are performed.}
We wish to store ${\bf f}$ across $k$ systematic and $2$ parity storage units each having storage capacity $\frac{M}{k}=N$, hence we consider a data rate of $\frac{k}{k+2}$.
We require that the encoded storage array is resilient up to any $2$ node erasures.
To satisfy the redundancy and erasure resiliency properties, the file is encoded using a $(k+2,k)$ MDS distributed storage code.
A storage code has the MDS property when any possible collections of $k$ storage nodes can reconstruct the file ${\bf f}$.

\begin{figure}[h]
\begin{align}
&\begin{array}{|c|c|}
\hline
\text{systematic node} & \text{systematic data}\\
\hline
1&{\bf f}_1\\
\hline
\vdots&\vdots\\
\hline
k&{\bf f}_k\\
\hline
\text{parity node} & \text{parity data}\\
\hline
1&{\bf f}_1+\ldots+{\bf f}_k\\
\hline
2&{\bf A}_1^T{\bf f}_1+\ldots+{\bf A}_k^T{\bf f}_k\\
\hline
\end{array}\nonumber
\end{align}
\caption{\textsc{A $(k+2,k)$ Coded Storage Array.}}
\end{figure}

In Fig. 1 we provide a general structure of  a two parity MDS encoded storage array.
The first $k$ storage nodes store the systematic file parts.
Without loss of generality, the first parity stores the sum of all $k$ systematic parts ${\bf f}_1+\ldots+{\bf f}_k$ and the second parity stores a linear combination of them ${\bf A}^T_1{\bf f}_1+\ldots+{\bf A}^T_k{\bf f}_k$.
Here, ${\bf A}_i$ denotes an $N\times N$ matrix of coding coefficients used by the second parity node  to ``scale and mix'' the contents of the $i$th file piece ${\bf f}_{i}$, $i\in\{1,\ldots,k\}$.
This representation is a  systematic one: $k$ nodes store uncoded file pieces and each of the $2$ parities stores a linear combination of the $k$ file parts.

In this work, we are interested in maintaining the same level of redundancy when a storage component fails or leaves the system.
To do that the {\it code repair} process has to take place to exactly regenerate the lost data in a {\it newcomer} storage component.
Let, for example, a systematic node $i\in\{1,\ldots,k\}$ fail.
Then, a newcomer joins the storage network, connects to the remaining nodes, and has to download sufficient data to reconstruct ${\bf f}_i$.

It is important to note that the lost systematic part ${\bf f}_i$, exists {\it only} as a term of a linear combination  at each parity node, as seen in Fig. 1.
Therefore, to regenerate the $N$ elements of ${\bf f}_i$, the newcomer has to download from the parity nodes a size of data equal to the size of the lost piece, i.e., $N$ linearly independent coded elements.
Assuming that it downloads the same amount of data from both parities, the downloaded contents can be represented as a stack of $N$ equations
{\small
\begin{align}
\hspace{-0.1cm}\left[
\begin{array}{c}
{\bf p}_i^{(1)}\\
{\bf p}_i^{(2)}
\end{array}
\right]\hspace{-0.1cm}&\eqdef\hspace{-0.1cm}
\left[
\begin{array}{c}
\left({\bf V}_i^{(1)}\right)^T{\bf f}_1+\ldots +\left({\bf V}_i^{(1)}\right)^T{\bf f}_k\\ 
\left({\bf V}_i^{(2)}\right)^T{\bf A}_1^T{\bf f}_1+\ldots +\left({\bf V}_i^{(2)}\right)^T{\bf A}_k^T{\bf f}_k 
\end{array}
\right]=\underbrace{\left[
\begin{array}{@{}c@{}}
\left({\bf V}^{(1)}_i\right)^T\\
\left({\bf A}_i{\bf V}^{(2)}_i\right)^T
\end{array}
\right]{\bf f}_i}_{\text{useful data}}+\sum_{s=1,s\ne i}^k\hspace{-0.1cm}
\underbrace{\left[
\begin{array}{@{}c@{}}
\left({\bf V}^{(1)}_i\right)^T\\
\left({\bf A}_s{\bf V}^{(2)}_i\right)^T\label{Yrep}
\end{array}
\right]\hspace{-0.1cm}{\bf f}_s}_{\text{interference by ${\bf f}_s$}},
\end{align}
}where ${\bf p}_i^{(1)},{\bf p}_i^{(2)}\in\mathbb{F}_q^{\frac{N}{2}}$ are the equations downloaded from the first and second parity node, respectively, and ${\bf V}_i^{(1)},{\bf V}_i^{(2)}\in\mathbb{F}_q^{N\times \frac{N}{2}}$ are the {\it repair matrices}.
Each repair matrix is used to mix the $N$ parity contents so that a set of $\frac{N}{2}$ equations is formed.
Then,  retrieving ${\bf f}_i$ from (\ref{Yrep}) is equivalent to solving an underdetermined set of $N$ equations in the $kN$ unknowns of ${\bf f}$, with respect to the $N$ desired unknowns of ${\bf f}_i$.
However, this is not possible due to $k-1$ additive {\it interference} components in the received equations generated by the undesired unknowns ${\bf f}_s$, $s\in\{1,\ldots,k\}\backslash i$, as noted in (\ref{Yrep}).
These $k-1$ interference terms corrupt the desired data and need to be canceled.
Hence, the newcomer needs to download additional data from the remaining $k-1$ systematic nodes, that will ``replicate'' and cancel the interference terms from the downloaded equations.

\begin{figure}[t]
\centerline{\includegraphics[width=0.5\columnwidth]{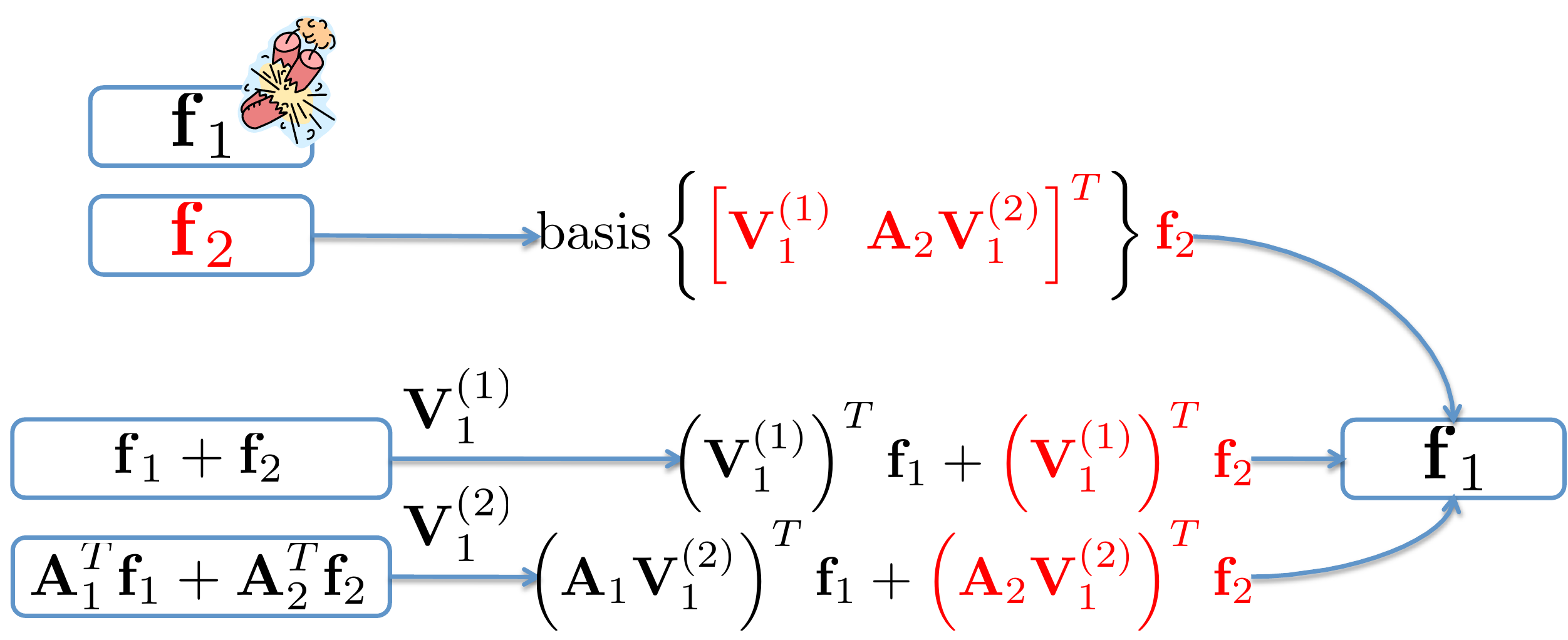}}
\caption{Repair of a $(4,2)$ code. Let systematic node $1$ fail. Then, a newcomer node joins the system and downloads data from the $3$ remaining nodes to regenerate ${\bf f}_1$. The useful information is mixed with the undesired part ${\bf f}_2$ in both information chunks downloaded from the parities. These interference parts are highlighted in red. To retrieve ${\bf f}_1$ a basis of the interference equations needs to be downloaded by systematic node $2$. Then, the newcomer can erase interference and invert the matrix multiplying ${\bf f}_1$ to retrieve it. Note that for invertibility, we need the additional condition that the matrix $\left[\mathbf{V}_{1}^{(1)}\;\;\;\mathbf{A}_{1}\mathbf{V}_{1}^{(2)}\right]^{T}$ has a full rank of $N$.
}
\end{figure}

To cancel a single interference term of (\ref{Yrep})  that has size $N$, it suffices to download a basis of equations that generates it.
The dimensions of this basis does not need to be equal to $N$. 
For example, to erase 
\begin{equation}
\left[
\begin{smallmatrix}
\left({\bf V}^{(1)}_i\right)^T\\
\left({\bf A}_s{\bf V}^{(2)}_i\right)^T
\end{smallmatrix}
\right]\hspace{-0.1cm}{\bf f}_s, \; s\in\{1,\ldots,k\}\backslash i \label{interfterm}
\end{equation}
the newcomer needs to connect to systematic node $s$ and download a number of linear equations in ${\bf f}_s$ that can generate (\ref{interfterm}); this number is equal to 
\begin{equation}
\frac{N}{2} \le \rank\left(\left[
\begin{smallmatrix}
\left({\bf V}^{(1)}_i\right)^T\\
\left({\bf A}_s{\bf V}^{(2)}_i\right)^T
\end{smallmatrix}
\right]\right)\le N, \label{bound}
\end{equation}
This is exactly the communication bandwidth price we are paying to delete a single interference term in order to be able to reconstruct ${\bf f}_i$. 
The lower bound in (\ref{bound}) comes from the fact that $\frac{N}{2}$ linearly independent equations need to be downloaded from each of the parities, hence $\rank({\bf V}_i^{(1)})=\rank({\bf V}_i^{(2)})=\frac{N}{2}$ for any $i\in\{1,\ldots,k\}$.
Eventually, we need to generate all undesired terms in the newcomer, so to subtract them from (\ref{Yrep}).
Then, a full rank system of $N$ equations in the $N$ unknowns has to be formed.
A generic example of a code repair instance for  a $(4,2)$ storage code is given in Fig. 2.

In general, to repair a systematic node $i\in\{1,\ldots,k\}$ of an arbitrary $(k+2,k)$ MDS storage code, we need to obtain a feasible solution to the following rank constrained, rank minimization (performed over $\mathbb{F}_q$)
\begin{equation}
\begin{split}
{\mathcal R}_i\text{: }
&\min_{{\bf V}^{(1)}_i,{\bf V}^{(2)}_i}\sum_{s=1,s\ne i }^k\rank\left(\left[{\bf V}^{(1)}_i \;\; {\bf A}_s{\bf V}^{(2)}_i\right]\right)\\
&\text{s.t.: }\rank\left(\left[{\bf V}^{(1)}_i \;\; {\bf A}_i{\bf V}^{(2)}_i\right]\right)=N,
\end{split}
\nonumber
\end{equation}
where {\it i)} the full rank constraints correspond to the requirement that the $N$ equations downloaded from the parities are linearly independent, when viewed as equations in the $N$ components of $\mathbf{f}_{i}$ and {\it ii)} the rank minimization corresponds to minimizing the sum of bases dimensions needed to cancel each interference term.
For a specific feasible selection of repair matrices the {\it repair bandwidth} to exactly regenerate systematic node $i$ is given by
{\small
\begin{align}
\gamma_i&=\underbrace{N}_{\text{\#equations lost}}+\sum_{s=1,s\ne i}^{k}\underbrace{\rank\left(\left[{\bf V}^{(1)}_i \;\;\;{\bf A}_s{\bf V}^{(2)}_i\right]\right)}_{\text{dim. of interference equations by ${\bf f}_s$}}=N+\sum_{s=1,s\ne i}^{k}\rank\left(\left[{\bf V}^{(1)}_i \;\;\;{\bf A}_s{\bf V}^{(2)}_i\right]\right),
\end{align}
}where the sum rank term is the aggregate of interference dimensions.
An optimal solution to  $\mathcal{R}_i$ is guaranteed to minimize the repair bandwidth we need to communicate to the repair systematic node $i\in\{1,\ldots,k\}$.

\begin{figure*}[t!]

{\footnotesize
\begin{align}
{\bf A}_1 = \diag\left(
9\left[
\begin{smallmatrix}
 1\\
     1\\
     1\\
     1\\
     1\\
     1\\
     1\\
     1\\
    -1\\
    -1\\
    -1\\
    -1\\
    -1\\
    -1\\
    -1\\
    -1
\end{smallmatrix}
\right]
+
7\left[
\begin{smallmatrix}
1\\
    -1\\
     1\\
    -1\\
     1\\
    -1\\
     1\\
    -1\\
     1\\
    -1\\
     1\\
    -1\\
     1\\
    -1\\
     1\\
    -1
\end{smallmatrix}
\right]
+
\left[
\begin{smallmatrix}
1\\
1\\
1\\
1\\
1\\
1\\
1\\
1\\
1\\
1\\
1\\
1\\
1\\
1\\
1\\
1
\end{smallmatrix}
\right]
\right),
\;\;
{\bf A}_2 = \diag\left(
5\left[
\begin{smallmatrix}
 1\\
     1\\
     1\\
     1\\
    -1\\
    -1\\
    -1\\
    -1\\
     1\\
     1\\
     1\\
     1\\
    -1\\
    -1\\
    -1\\
    -1
\end{smallmatrix}
\right]
+
2\left[
\begin{smallmatrix}
1\\
    -1\\
     1\\
    -1\\
     1\\
    -1\\
     1\\
    -1\\
     1\\
    -1\\
     1\\
    -1\\
     1\\
    -1\\
     1\\
    -1
\end{smallmatrix}
\right]
+
\left[
\begin{smallmatrix}
1\\
1\\
1\\
1\\
1\\
1\\
1\\
1\\
1\\
1\\
1\\
1\\
1\\
1\\
1\\
1
\end{smallmatrix}
\right]
\right),
\;\;
{\bf A}_3 = \diag\left(
9\left[
\begin{smallmatrix}
 1\\
     1\\
    -1\\
    -1\\
     1\\
     1\\
    -1\\
    -1\\
     1\\
     1\\
    -1\\
    -1\\
     1\\
     1\\
    -1\\
    -1
\end{smallmatrix}
\right]
+
4\left[
\begin{smallmatrix}
1\\
    -1\\
     1\\
    -1\\
     1\\
    -1\\
     1\\
    -1\\
     1\\
    -1\\
     1\\
    -1\\
     1\\
    -1\\
     1\\
    -1
\end{smallmatrix}
\right]
+
\left[
\begin{smallmatrix}
1\\
1\\
1\\
1\\
1\\
1\\
1\\
1\\
1\\
1\\
1\\
1\\
1\\
1\\
1\\
1
\end{smallmatrix}
\right]
\right)\nonumber 
\end{align}
}
\hrulefill
\caption{A repair optimal $(5,3)$ MDS code over $\mathbb{F}_{11}$}

\end{figure*}

\begin{rem}
From \cite{DimakisGWWR:08}  it is known that the theoretical minimum repair bandwidth, for any single node repair of an optimal (linear or nonlinear) $(k+2,k)$ MDS code is exactly $(k+1)\frac{N}{2}$, where $N$ has to be an even number.
This bound is proven using cut-set bounds on infinite flow graphs.
Here, we provide an interpretation of this bound in terms of linear codes by calculating the minimum possible sum of ranks in $\mathcal{R}_i$: since each repair matrix {\it has} to have full column rank $\frac{N}{2}$ to be a feasible solution, the minimum number of dimensions each interference can be suppressed to is $\frac{N}{2}$. This aggregates in a minimum repair bandwidth of $(k+1)\frac{N}{2}$ repair equations.
If we wish to achieve this bound, {\it interference alignment} has to be employed, so that undesired components like (\ref{interfterm}) are confined to the minimum number of dimensions.
Interestignly, linear codes suffice to asymptotically achieve this bound \cite{CadambeWinc}, \cite{SuhCodes}.
\end{rem}

We know what the theoretical minimum repair bandwidth is and that there exist asymptotically optimal schemes, however, designing MDS codes with repair strategies that achieve it has been challenging.
The difficulty in designing optimal MDS storage codes lies in a threefold requirement: {\it i)} the code has to satisfy the MDS property, {\it ii)} systematic nodes of the code have to be optimally repaired, and {\it iii)}  parity nodes of the code have to be optimally repaired.
Currently, there exist MDS codes for rates $\frac{k}{n}\le \frac{1}{2}$ \cite{RashmiProduct, SuhR:09} for which all nodes can be optimally repaired.
For the high data rate regime, Tamo {\it et al.} \cite{Tamo} and Cadambe {\it et al.} \cite{PermCodes} presented the first MDS codes where any systematic node failure can be optimally repaired.
However, prior to this work, there do not exist MDS storage codes of arbitrarily high rate that can optimal repair {\it any} node.

In the following, we present the first explicit, high-rate, repair optimal $(k+2,k)$ MDS storage code that achieves the minimum repair bandwidth bound for the repair of {\it any} single systematic or parity node failure.

\section{A Repair Optimal $2$ Parity Storage Code}
Let a $(k+2,k)$ MDS storage code for file size $M= k2^{k+1}$, with coding matrices
\begin{equation}
{\bf A}_i = a_i{\bf X}_i+b_i{\bf X}_{k+1}+{\bf I}_{N},\; i\in\{1,\ldots,k\}\label{code}
\end{equation}
where $N=2^{k+1}$,
\begin{equation}
{\bf X}_i = {\bf I}_{2^{i-1}}\otimes \text{blkdiag}\left({\bf I}_{\frac{N}{2^{i}}},-{\bf I}_{\frac{N}{2^{i}}}\right),
\end{equation}
and $a_i$, $b_i$ satisfy $a_i^2-b_i^2=-1$, for all $i\in\{1,\ldots,k\}$.\footnote{We use $-1$ to denote the field element $q-1$ over $\mathbb{F}_q$.}

\begin{theo}
There exists a finite field $\mathbb{F}_q$ of order $q\ge 2k+3$ and explicit constants $a_i,b_i\in\mathbb{F}_q,\,\forall i\in\{1,\ldots,k\}$, such that the $(k,k+2)$ storage code in (\ref{code}) is a repair optimal MDS storage code.\label{1}
\end{theo}
In Fig. 3, we give the coding matrices of a $(5,3)$ MDS code over $\mathbb{F}_{11}$ based on our construction.

\begin{rem}
The code constructions presented here have generator matrices that are as sparse as possible, since any additional sparsity would violate the MDS property.
This creates the additional benefit of minimum update complexity when some bits of the stored data object change.
\end{rem}

Before we proceed with proving Theorem \ref{1}, we state the intuition behind our code construction and the tools that we use.
Motivated by the asymptotic IA schemes, we use similar concepts motivated by a combinatorial explanation of interference alignment in terms of dots on lattices. 
In contrast to the asymptotic IA codes, here, instead of letting randomness choose the coding matrices, we select particular constructions based on  Hadamard matrices that achieve {\it exact} interference alignment for fixed in $k$ file sizes (symbol extensions).
In section V we prove the optimal repair of systematic nodes, in Section VII we show the optimal repair of parity nodes, and in section VIII we state explicit conditions for the MDS property.

\section{Dots-on-a-lattice and Hadamard Designs}
In Section II, we showed that minimizing the communication bandwidth to repair nodes of a storage code is equivalent to the problem of minimizing the dimensions of interference terms generated during each repair process.
Here, we consider the problem of designing coding and repair matrices that can achieve {\it perfect} interference alignment in a finite number of extensions.
We begin by assuming arbitrary constructions and then we use a combinatorial explanation of IA to find conditions under which perfect alignment in the finite file regime.
Eventually, we show that exact IA conditions and linear independence requirements posed by our problem are simultaneously satisfied through the use of Hadamard designs.

Assume two arbitrary $N\times N$ full rank matrices ${\bf T}_1$ and ${\bf T}_2$ that commute.
We wish to construct a full rank matrix ${\bf V}$, with at most $\frac{N}{2}$ columns, such that the span of ${\bf T}_1{\bf V}$ aligns as much as possible with the span of ${\bf T}_2{\bf V}$:  we have to pick ${\bf V}$ such that it minimizes the dimensions of the union of the two spans, that is the rank of $\left[{\bf T}_1{\bf V} \;\;\;{\bf T}_2{\bf V}\right]$.
How can we construct such a matrix?
Assume that we start with one vector with nonzero entries, i.e., ${\bf V}={\bf w}$, and for simplicity we let it be the all-ones vector.
Then in the general case, ${\bf T}_1{\bf w}$ and ${\bf T}_2{\bf w}$ have zero intersection which is not desired.
However, we can augment ${\bf V}$ such that it has as columns the elements of the set $\{{\bf w},{\bf T}_1{\bf w},{\bf T}_2{\bf w},{\bf T}_1{\bf T}_2{\bf w}\}$.
Observe that each vector ${\bf T}^{x_1}_1{\bf T}^{x_2}_2{\bf w}$ of ${\bf V}$ can be represented by the power tuple $(x_1,x_2)$.
This helps us visualize ${\bf V}$ as a set of dots on the $2$-dimensional integer lattice as shown in Fig. 4.
\begin{figure}[h]
\centerline{\includegraphics[width=0.2\columnwidth]{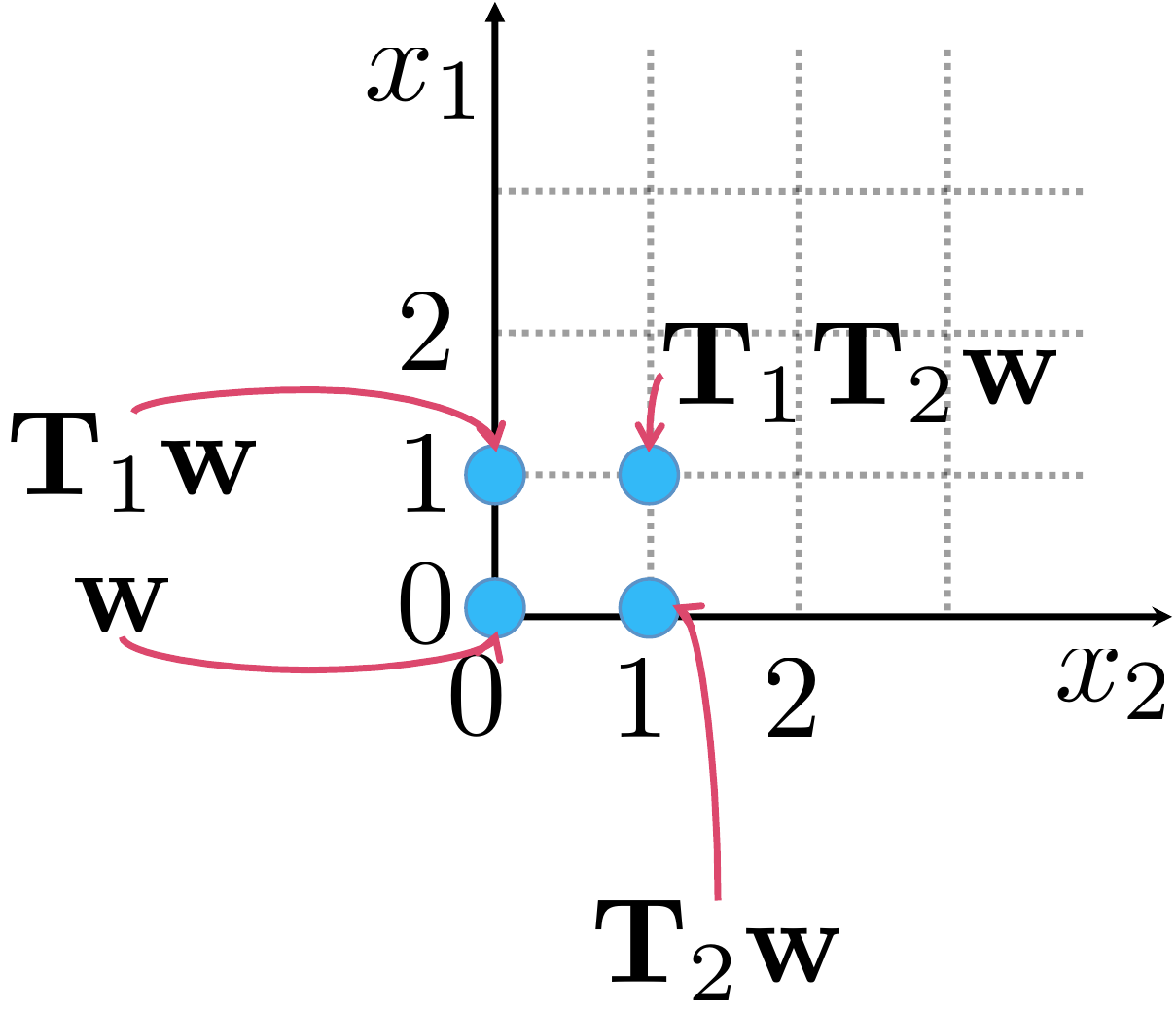}}
\caption{Representing ${\bf V}$ as dots on a lattice.}
\end{figure}
\\
For this new selection of ${\bf V}$, we have 
\begin{align}
{\bf T}_1{\bf V}& = \left[{\bf T}_1{\bf w}\;\;\;{\bf T}^2_1{\bf w}\;\;\;{\bf T}_1{\bf T}_2{\bf w}\;\;\;{\bf T}^2_1{\bf T}_2{\bf V}\right]\\
\text{and }{\bf T}_2{\bf V}& = \left[{\bf T}_2{\bf w}\;\;\;{\bf T}_2{\bf T}_1{\bf w}\;\;\;{\bf T}^2_2{\bf w}\;\;\;{\bf T}_1{\bf T}^2_2{\bf V}\right].
\end{align}
The intersection of the spans of these two matrices is now nonzero: the matrix $\left[{\bf T}_1{\bf V} \;\;\;{\bf T}_2{\bf V}\right]$ has rank $7$ instead of the maximum possible of $8$. 
This happens because the vector ${\bf T}_1{\bf T}_2{\bf w}$ is repeated in both matrices ${\bf T}_1{\bf V}$ and ${\bf T}_2{\bf V}$.
In Fig. 5 we illustrate this concatenation, in terms of dots on $\mathbb{Z}^2$, where the intersection between the two spans is manifested as an overlap of dots.
\begin{figure}[h]
\centerline{\includegraphics[width=0.2\columnwidth]{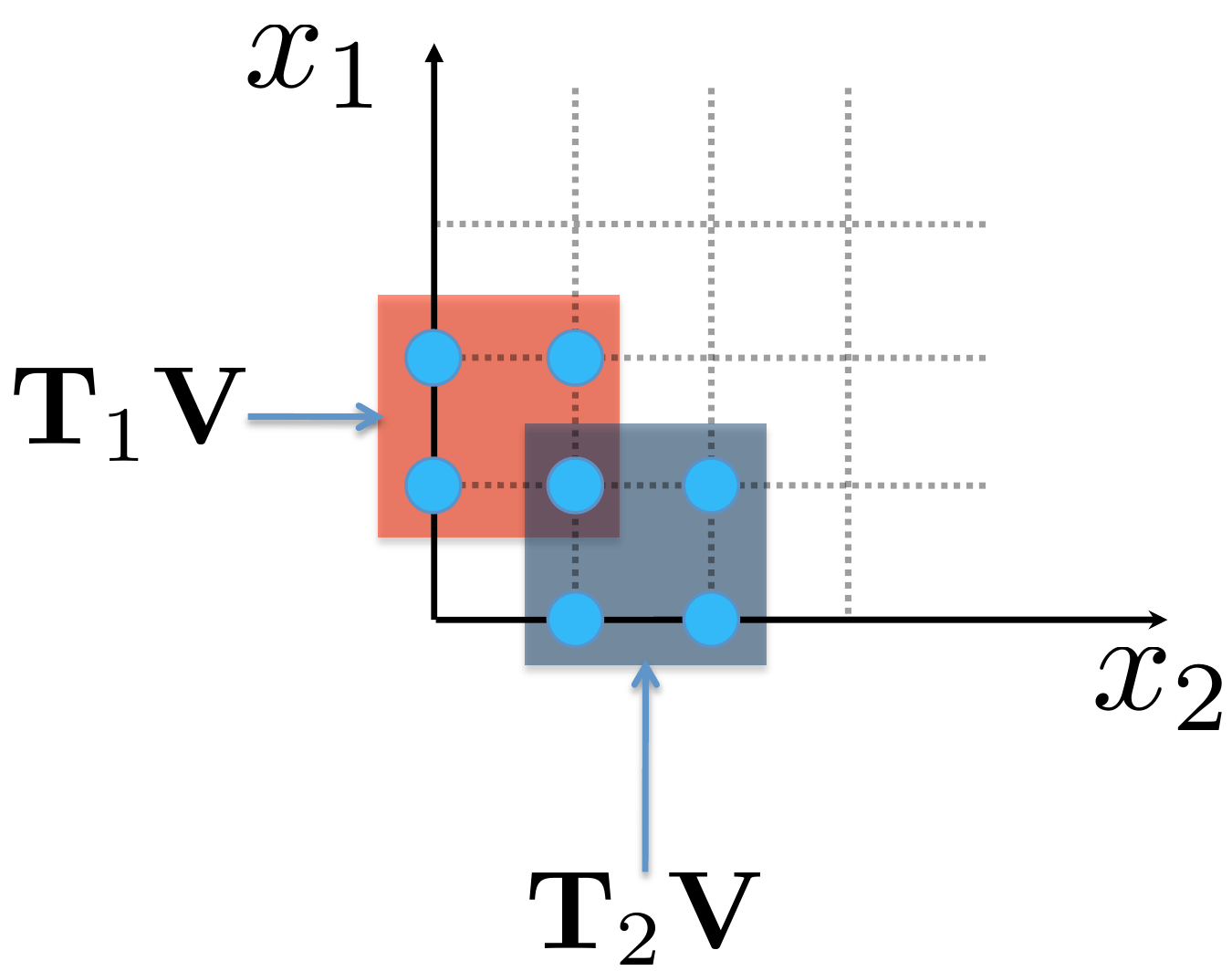}}
\caption{Representing $\left[{\bf T}_1{\bf V} \;\;\;{\bf T}_2{\bf V}\right]$ as dots on a lattice.}
\end{figure}
\\
\begin{rem}
Observe how matrix multiplication of ${\bf T}_1$ and ${\bf T}_2$ with the vectors in ${\bf V}$ is pronounced through the dots representation: the dots representations of ${\bf T}_1{\bf V}$ and ${\bf T}_2{\bf V}$ matrices are shifted versions of ${\bf V}$ along the $x_1$ and $x_2$ axes.
\end{rem}
The key idea behind choosing a new ${\bf V}$ at each step is to iteratively augment the old one with products of the ${\bf T}_i$ matrices raised to specific powers times the current ${\bf V}$ 
{\small
\begin{align}
\text{\texttt{initialize }:\;\; }{\bf V} &\leftarrow {\bf w}\\
\text{\texttt{multiply with powers of } ${\bf T}_1$:\;\; }{\bf V} &\leftarrow \left[ {\bf V} \;\;\;{\bf T}_1{\bf V} \ldots\;\;\;{\bf T}_1^{m-1}{\bf V}\right]\\
\text{\texttt{multiply with powers of } ${\bf T}_2$: \;\;}{\bf V} &\leftarrow \left[ {\bf V} \;\;\;{\bf T}_2{\bf V} \ldots\;\;\;{\bf T}_2^{m-1}{\bf V}\right].
\end{align}
}

In general, by using powers up to $m$, with $m^2\le \frac{N}{2}$,  we obtain ${\bf V}$ with $m^2$ columns that are the elements of the set
{\small
\begin{equation}
\mathcal{V}=\left\{{\bf T}_1^{x_1}{\bf T}_2^{x_1}{\bf w}:x_s\in\{0,\ldots,m-1\}\right\},
\end{equation}
}
where ${\bf w}={\bf 1}_{N\times 1}$.
Then, matrix ${\bf V}$ achieves the following property
\begin{align}
m^2<\rank\left(\left[{\bf T}_1{\bf V}\;\;\;{\bf T}_2{\bf V}\right]\right) < (m+1)^2,
\end{align}
which means that we can asymptotically create as much alignment as we desire within the spans of the matrices ${\bf T}_i{\bf V}$, for arbitrarily large ``symbol extensions'', i.e. for sufficiently large $N$,  $(m+1)^2/m^2$ is arbitrarily close to $1$.
For example, we give the $m=4$ case in Fig. 6, where we observe that the alignment is more substantial (with respect to the size of ${\bf V}$) compared to Fig. 5.
\begin{figure}[h]
\centerline{\includegraphics[width=0.2\columnwidth]{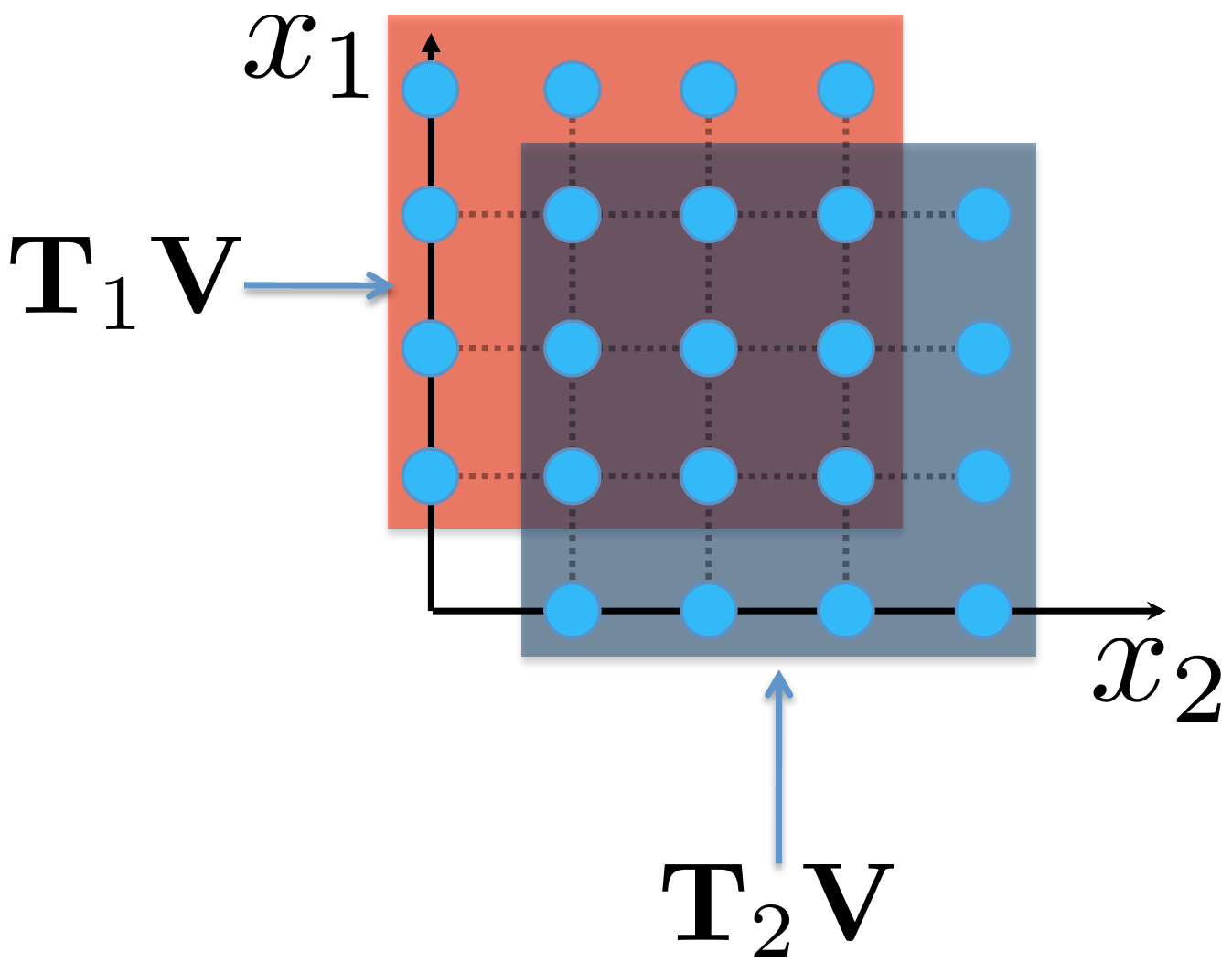}}
\caption{Representing $\left[{\bf T}_1{\bf V} \;\;\;{\bf T}_2{\bf V}\right]$ as dots on a lattice.}
\end{figure}
This alignment scheme, in a more general form, was presented by Cadambe and Jafar in \cite{CadambeJ:08} to prove the Degrees-of-Freedom of the $K$-user interference channel.
For that wireless scenario, the ${\bf T}_i$ matrices are given by nature and are i.i.d. diagonals. 
Perfect alignment of spaces for these matrices is not known to be possible for finite $m$ \cite{CadambeJ:08, Bresler_Tse_diversity}.

For network coding problems, and in particular, for storage coding problems, the analogous ${\bf T}_i$ matrices (our coding matrices) are free to design under some specific constraints that ensure the MDS property of the code.
Before, we give explicit matrices that achieve alignment in a finite number of extensions, we answer the analogous question considering our toy example: do there exist ${\bf T}_1$ and ${\bf T}_2$ matrices such that we can construct a full-rank ${\bf V}$ that achieves perfect intersection (exact alignment) of the spans of ${\bf T}_1{\bf V}$ and ${\bf T}_2{\bf V}$, for some $m$ and $N= m^3$?
That is, can we find matrices such that
\begin{align}
\Span({\bf T}_1{\bf V})=\Span({\bf T}_2{\bf V}) \text{ and } \rank({\bf V})=m^2
\end{align}
is possible?
We show that a sufficient condition for perfect alignment is satisfied when the elements of the matrices are $m^{\text{th}}$ roots of unity, i.e.,
\begin{equation}
{\bf T}_i^m = {\bf I}_N.
\end{equation}
To see, that we formally state the dots on a lattice representation.
Let a map $\mathcal{L}$ from a matrix with $r$ columns, each generated as ${\bf T}^{x_1}_1{\bf T}^{x_2}_2{\bf w}$, to a set of $r$ points, such that
the column ${\bf T}^{x_1}_1{\bf T}^{x_2}_2{\bf w}$ maps to the point $(x_1,x_2)$.
Then, we have for ${\bf V}$
\begin{equation}
\mathcal{L}({\bf V})\eqdef\left\{x_1{\bf e}_1+x_2{\bf e}_2;\; x_1,x_2\in[m]\right\},
\end{equation}
where $[m] = \{0,\ldots,m-1\}$ and ${\bf e}_i$ is the $i$-th column of the identity matrix.
Using this representation, the products ${\bf T}_1{\bf V}$ and ${\bf T}_2{\bf V}$ map to 
{\small
\begin{align}
\mathcal{L}({\bf T}_1{\bf V})=\Biggl\{\hspace{-0.1cm}(x_1+1){\bf e}_1+x_2{\bf e}_2:x_1,x_2\in[m]\hspace{-0.1cm}\Biggr\}\text{ and }\mathcal{L}({\bf T}_2{\bf V})=\Biggl\{\hspace{-0.1cm}x_1{\bf e}_1+(x_2+1){\bf e}_2:x_1,x_2\in[m]\hspace{-0.1cm}\Biggr\}
\end{align}
}respectively.
For perfect alignment, we have to design the ${\bf T}_i$ matrices such that
\begin{equation}
\mathcal{L}({\bf T}_1{\bf V})=\mathcal{L}({\bf T}_2{\bf V}).
\end{equation}
A sufficient set of conditions for perfect span instersection is that ${\bf V}$, ${\bf T}_1{\bf V}$, and ${\bf T}_2{\bf V}$ perfectly intersect, i.e.
{\small
\begin{align}
\mathcal{L}({\bf T}_1{\bf V})=\mathcal{L}({\bf V})
&\Leftrightarrow\Biggl\{\hspace{-0.1cm}(x_1+1){\bf e}_1+x_2{\bf e}_2: x_1,x_2\in[m]\Biggr\}
=\Biggl\{\hspace{-0.1cm}x_1{\bf e}_1+x_2{\bf e}_2: x_1,x_2\in[m]\Biggr\},\\
\mathcal{L}({\bf T}_1{\bf V})=\mathcal{L}({\bf V})
&\Leftrightarrow\Biggl\{\hspace{-0.1cm}x_1{\bf e}_1+(x_2+1){\bf e}_2: x_1,x_2\in[m]\Biggr\}
=\Biggl\{\hspace{-0.1cm}x_1{\bf e}_1+x_2{\bf e}_2: x_1,x_2\in[m]\Biggr\}.
\end{align}
}The above conditions are satisfied when the matrix powers ``wrap around'' upon reaching a certain modulus $m$.
This wrap-around property is obtained when the ${\bf T}_1$ and ${\bf T}_2$ matrices have elements that are $m^{\text{th}}$ roots of unity
\begin{equation}
{\bf T}_1^m = {\bf T}_1^0={\bf T}_2^m = {\bf T}_2^0={\bf I}_N.
\end{equation}
However, arbitrary diagonal matrices whose elements are $m^{\text{th}}$ roots of unity are not sufficient to ensure the full rank property of ${\bf V}$.
To hint on a general procedure which outputs ``good'' ${\bf T}_i$ matrices, we see an example where we pick them such that ${\bf V}$ has orthogonal columns. 
Let us briefly consider the case where $m=2$ and $N = 2^3$, for which we choose
{\small
\begin{equation}
{\bf T}_1 = \diag\left(
\left[
\begin{smallmatrix}
1\\
-1\\
1\\
-1\\
1\\
-1\\
1\\
-1\\
\end{smallmatrix}
\right]
\right)
\;\text{ and }\;{\bf T}_2 =\diag\left(
\left[
\begin{smallmatrix}
1\\
1\\
-1\\
-1\\
1\\
1\\
-1\\
-1\\
\end{smallmatrix}
\right]\right).
\label{eq:X1X2}
\end{equation}
}For these matrices, ${\bf V}$ has $m^2 = 4$ orthogonal columns
{\small
\begin{equation}
{\bf V} = \left[{\bf w}\;\;\;{\bf T}_1{\bf w}\;\;\;{\bf T}_2{\bf w}\;\;\;{\bf T}_1{\bf T}_2{\bf w}\right]=
\left[
\begin{smallmatrix}
1&1&1&1\\
1&-1&1&-1\\
1&1&-1&-1\\
1&-1&-1&1\\
1&1&1&1\\
1&-1&1&-1\\
1&1&-1&-1\\
1&-1&-1&1
\end{smallmatrix}
\right]
\label{eq:V}
\end{equation}
}and ${\bf T}_2{\bf V} = \left[{\bf T}_2{\bf w}\;\;\;{\bf T}_1{\bf T}_2{\bf w}\;\;\;{\bf w}\;\;\;{\bf T}_1{\bf w}\right] $, ${\bf T}_3{\bf V} = \left[{\bf T}_1{\bf w}\;\;\;{\bf w}\;\;\;{\bf T}_1{\bf T}_2{\bf w}\;\;\;{\bf T}_2{\bf w}\right] $ indeed have fully overlapping spans.
Interestingly, we observe that for the additional matrix 
{\small
\begin{equation}
{\bf T}_3 = \diag\left(
\left[
\begin{smallmatrix}
1\\
1\\
1\\
1\\
-1\\
-1\\
-1\\
-1\\
\end{smallmatrix}
\right]\right)
\end{equation}
}we have that $\left[{\bf V}\;\;\;{\bf T}_3{\bf V}\right] = {\bf H}_8$,
where ${\bf H}_8$ is the $8\times 8$ Hadamard matrix.
In the following we see that Hadamard designs provide the conditions for perfect alignment and linear independence in a more general setting.

Let $m=2$, $N=2^L$, and ${\bf X}_i = {\bf I}_{2^{i-1}}\otimes \text{blkdiag}\left({\bf I}_{\frac{N}{2^{i}}},-{\bf I}_{\frac{N}{2^{i}}}\right)$, for $i\in[L]$, and consider the set
\begin{equation}
\mathcal{H}_{N} = \left\{\prod_{i = 1}^{L}{\bf X}_{i}^{x_i}{\bf w}: x_{i}\in \{0,1\}\right\}. \label{Hprod}
\end{equation}
\begin{lem}
Let  an $N\times N$ Hadamard matrix of the Sylvester's construction
{\small
\begin{equation}
{\bf H}_{N} \eqdef \left[
\begin{array}{rr}
{\bf H}_{\frac{N}{2}} &{\bf H}_{\frac{N}{2}}\\
{\bf H}_{\frac{N}{2}} & -{\bf H}_{\frac{N}{2}}
\end{array}
\right],
\end{equation}
}with ${\bf H}_{1} = 1$.
Then,
${\bf H}_{N}$ is full rank with mutually orthogonal columns, that are the $N$ elements of $\mathcal{H}_{N}$.
\label{HadamardLem}
\end{lem}
The proof of Lemma (\ref{HadamardLem}) can be found in the Appendix.
\begin{example}
To illustrate the connection between $\mathcal{H}_{N}$ and ${\bf H}_{N}$ we ``decompose'' the Hadamard matrix of order $4$
{\small
\begin{align}
{\bf H}_4 &= \left[
\begin{smallmatrix}
1 & 1 & 1 & 1\\
1 & -1 & 1 & -1\\
1 & 1 & -1 & -1\\
1 & -1 & -1 & 1
\end{smallmatrix}
\right] = \left[{\bf w}\;\;{\bf X}_2{\bf w}\;\;{\bf X}_1{\bf w}\;\;{\bf X}_2{\bf X}_1{\bf w}\right],
\end{align}
}where 
${\bf X}_1 = \text{diag}\left(
\begin{smallmatrix}
1\\
1\\
-1\\
-1
\end{smallmatrix}
\right) \text{ and } 
{\bf X}_2 = 
\diag\left(
\begin{smallmatrix}
1 \\
-1\\
1\\
-1
\end{smallmatrix}
\right)$.
Due to the commutativity of ${\bf X}_1$ and ${\bf X}_2$, the columns of ${\bf H}_{4}$ are also the elements of
$\mathcal{H}_{4}=\left\{{\bf w},{\bf X}_1{\bf w},{\bf X}_2{\bf w},{\bf X}_1{\bf X}_2{\bf w}\right\}$.
\end{example}
Now, consider the matrix ${\bf V}_i$ that has as columns the elements of
\begin{equation}
\mathcal{V}_i = \left\{\prod_{s=1,s\ne i}^{L} {\bf X}^{x_s}_s{\bf w}:x_s\in\{0,1\}\right\}. \label{Vi}
\end{equation} 
We know that the space of ${\bf V}_i$ is invariant with repsect to ${\bf X}_j$ since the corresponding lattice representation wraps around itself due to ${\bf X}^2_i={\bf I}_N$.
Additionally, we have
\begin{align}
\mathcal{L}({\bf X}_i{\bf V}_i)&=\left\{{\bf e}_i+\sum_{s=1,s\ne i}^{L}x_s{\bf e}_s: x_s\in\{0,1\}\right\},\nonumber
\end{align}
and we observe that $\mathcal{L}({\bf X}_i{\bf V}_i)\cap \mathcal{L}({\bf V}_i)=\emptyset$, i.e., $\mathcal{L}({\bf V}_i)$ does not include any points with nonzero $x_i$ coordinates.
Then, due to the orthogonality of elements within $\mathcal{H}_{N}$, we have
\begin{align}
|\mathcal{L}({\bf V}_i)|&=|\mathcal{L}({\bf X}_j{\bf V}_i)|=\rank({\bf V}_i)=\rank({\bf X}_i{\bf V}_i)=\frac{N}{2},
\end{align}
for any $i,j\in\{1,\ldots,L\}$.
Hence, we obtain the following lemma for the set $\mathcal{H}_{N}$ and its associated $\mathcal{L}$ map.
\begin{lem}
For any $i,j\in\{1,2,\ldots,L\}$ we have that
\begin{align}
\rank(\left[{\bf V}_i\;\;\;{\bf X}_j{\bf V}_i\right])&=\left|\mathcal{L}({\bf V}_i)\cup\mathcal{L}\left({\bf X}_j{\bf V}_i\right)\right|
=\left\{
\begin{array}{lc}
N, & i=j,\\
\frac{N}{2}, & i\ne j.
\end{array}
\right.
\end{align}
\end{lem}

In Fig. 7 we give an illustrative example of the aforementioned definitions and properties.
For $N = 2^3$, we consider ${\bf H}_{8}$ and ${\bf V}_3$ along with the matrix product ${\bf X}_2{\bf V}_3$ and their corresponding lattice representations.

\begin{figure}[h]
 \centerline{\includegraphics[width=0.4\columnwidth]{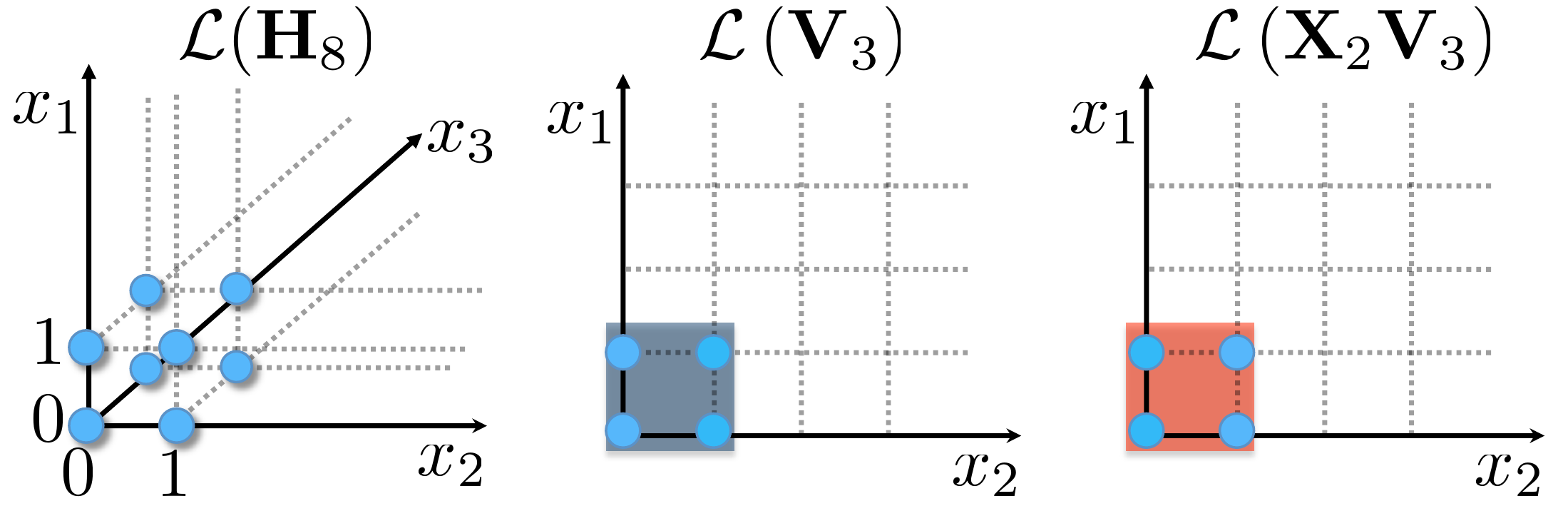}}
\caption{We set $N=8$ and show the dots representation of ${\bf H}_{8}$,  ${\bf V}_3$, and ${\bf X}_2{\bf V}_3$.}
\end{figure}

We use the aforementioned properties of Hadamard matrices to construct repair matrices ${\bf V}_i$ for our code construction; these matrices have perfect space alignment properties for the repair instances of the code in (\ref{code}) induced by single node failures.

\begin{rem}
Notice that equations (\ref{eq:X1X2}) and (\ref{eq:V}) are respectively analogous to the channel matrices and beamforming vectors used in wireless channels for ergodic interference alignment \cite{Nazer_Ergodic}. In particular, for the $K$ user interference channel, the channel matrices used for ergodic alignment are diagonalized versions of the column vectors of $\mathbf{H}_{2}.$
\end{rem}

\section{Optimal Systematic Node Repair}

Let systematic node $i\in\{1,\ldots,k\}$ of the code in (\ref{code}) fail.
The coding matrix ${\bf A}_i$ corresponding to the lost systematic piece ${\bf f}_i$, holds one matrix, that is, ${\bf X}_i$, which is unique among all other coding matrices, ${\bf A}_s$, $s\in\{1,\ldots,k\}\backslash i$. 
We pick the repair matrix as a set of $\frac{N}{2}$ vectors whose lattice representation is invariant to all ${\bf X}_j$s  but to one key matrix: the unique ${\bf X}_i$ component of ${\bf A}_i$. 
We construct the $N\times \frac{N}{2}$ repair matrix ${\bf V}_i$ whose columns are the elements of the set
\begin{equation}
\mathcal{V}_i = \left\{\prod_{s=1,s\ne i}^{k+1} {\bf X}^{x_s}_s{\bf w}:x_s\in\{0,1\}\right\}. \label{Vi}
\end{equation}
This repair matrix is used to multiply both the contents of parity node $1$ and $2$, that is, ${\bf V}_i^{(1)}={\bf V}_i^{(2)}={\bf V}_i$.
During the repair, the useful (desired signal) space populated by ${\bf f}_i$ is
\begin{equation}
\left[{\bf V}_i \;\; {\bf A}_i{\bf V}_i\right]
\end{equation}
and the interference space due to file part ${\bf f}_s$, $s\in\{1,\ldots,k\}\backslash i$, is 
\begin{equation}
\left[{\bf V}_i \;\; {\bf A}_s{\bf V}_i\right].
\end{equation}
Remember that an optimal solution to $\mathcal{R}_i$ requires the useful space to have rank $N$ and each of the interference spaces rank $\frac{N}{2}$.
Observe that the following holds for each of the interference spaces
\begin{align}
\frac{N}{2}&\le\rank\left(\left[{\bf V}_i \;\;\left(a_s{\bf X}_s+b_s{\bf X}_{k+1}+{\bf I}_{N}\right){\bf V}_i\right]\right)\nonumber\\
&\le\left|\mathcal{L}\left({\bf V}_i\right)\cup \mathcal{L}\left({\bf X}_s{\bf V}_i\right)\cup\mathcal{L}\left({\bf X}_{k+1}{\bf V}_i\right)\right|=\left|\mathcal{L}\left({\bf V}_i \right)\right|=\frac{N}{2},
\end{align}
for $s\in\{1,\ldots,k\}\backslash i$, since
\begin{equation}
\mathcal{L}({\bf X}_s{\bf V}_i)=\mathcal{L}({\bf V}_i), s\in\{1,\ldots,k+1\}\backslash i.
\end{equation}
Then, for the useful data space we have
\begin{align}
N &\ge\rank\left(\left[{\bf V}_i \;\;\;{\bf A}_i{\bf V}_i\right]\right)=\rank\left(\left[{\bf V}_i \;\;\left(a_i{\bf X}_i+b_i{\bf X}_{k+1}+{\bf I}_{N}\right){\bf V}_i\right]\right)\nonumber\\
&\overset{(*)}{=}\rank\left(\left[{\bf V}_i \;\;\;{\bf X}_i{\bf V}_i\right]\right) = \left|\mathcal{L}\left({\bf V}_i\right)\cup  \mathcal{L}\left({\bf X}_{i}{\bf V}_i\right)\right|\nonumber\\
&=\left|\mathcal{L}\left({\bf H}_{N}\right)\right|=N,
\end{align}
for any $a_i\ne 0$, where $(*)$ comes from the fact that $\left(a_i{\bf X}_i+b_i{\bf X}_{k+1}+{\bf I}_{N}\right){\bf V}_i$ is a linear combination of columns from ${\bf V}_i$, ${\bf X}_{k+1}{\bf V}_i$, and ${\bf X}_i{\bf V}_i$.
The column spaces of ${\bf V}_i$ and ${\bf X}_{k+1}{\bf V}_i$ are identical, hence we can generate the columns of $\left(a_i{\bf X}_i+b_i{\bf X}_{k+1}+{\bf I}_{N}\right){\bf V}_i$ by linear combinations of the columns in ${\bf X}_i{\bf V}_i$ and in ${\bf V}_i$, however ${\bf V}_i$ is already in the concatenation $\left[{\bf V}_i \;\;\left(a_i{\bf X}_i+b_i{\bf X}_{k+1}+{\bf I}_{N}\right){\bf V}_i\right]$. This means that $\left[{\bf V}_i \;\;\;{\bf X}_i{\bf V}_i\right]$ and $\left[{\bf V}_i \;\;\left(a_i{\bf X}_i+b_i{\bf X}_{k+1}+{\bf I}_{N}\right){\bf V}_i\right]$ have the same span.

Therefore, we are able to generate the minimum amount of interference and at the same time satisfiy the full rank constraint of $\mathcal{R}_i$.
The repair matrix in (\ref{Vi}) is an optimal solution for $\mathcal{R}_i$ and systematic node $i$ can be optimally repaired by downloading $(k+1)\frac{N}{2}$ worth data equations, for all $i\in\{1,\ldots,k\}$.
In Fig. 8, we  sketch the structure of our code. In each block of the second parity we denote the key matrices that comprise it.
We select our repair matrix such that it ``absorbs'' all matrices but the key one. 
That way, interference aligns in half the dimensions, and the useful space spans all $N$ dimensions.

\begin{figure*}[h]
\begin{center}
 \includegraphics[width=0.32\columnwidth]{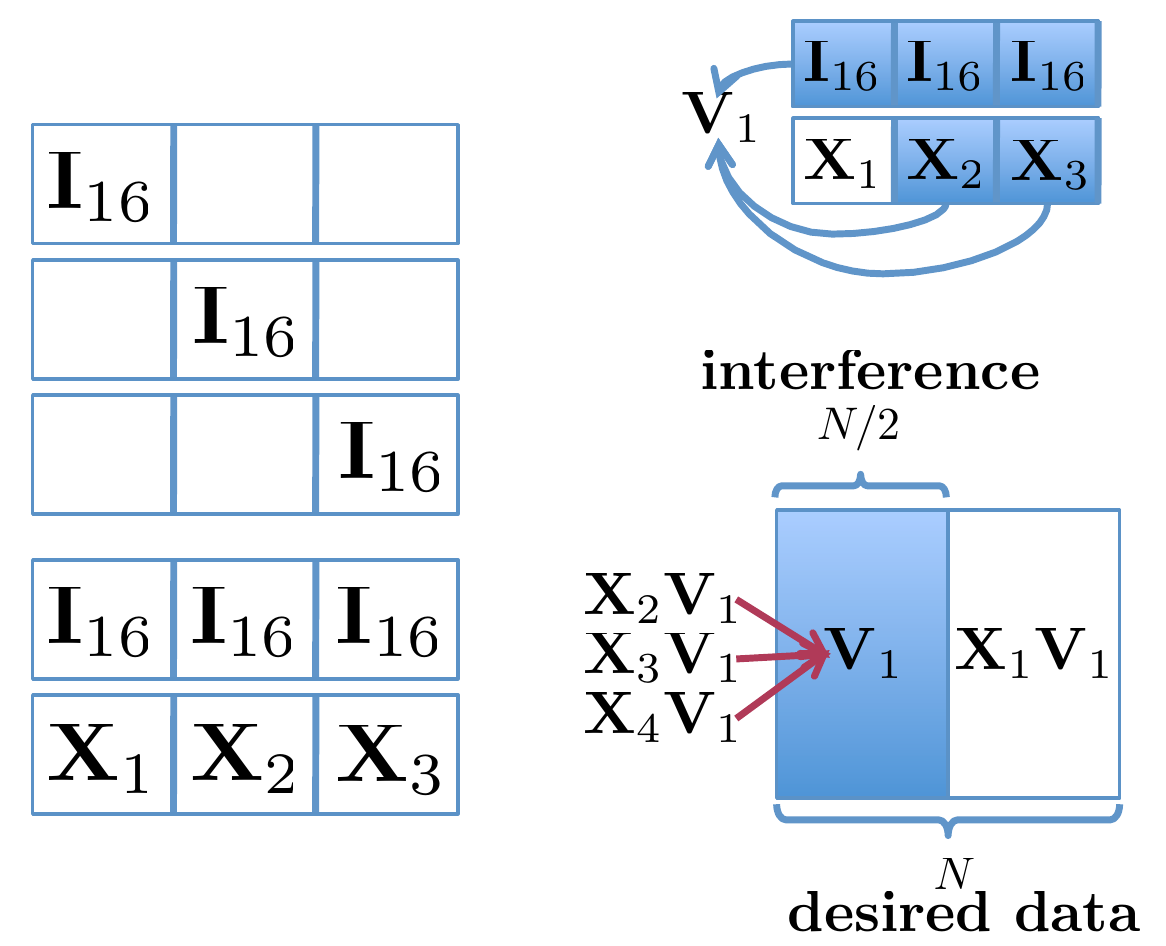}
\caption{A $(5,3)$ repair optimal code.}
\end{center}
\end{figure*}

\section{Optimal Parity Repair}

The ingredient of our construction that ``unlocks'' optimal repair for the first parity is the inclusion of the identity matrix in each ${\bf A}_i$.
The same goes for the ${\bf X}_{k+1}$ matrix and the repair of the second parity.
Both these additionally included matrices refine the parity repair process such that optimality is feasible.
Selecting appropriate constants $a_i$ and $b_i$ is also essential to our developments.
To optimally solve the problem, we rewrite the parity repair as a systematic one in an equivalent re-interpretation of our code.

\subsection{Repairing the first parity}
Let the first parity node fail.
We make a change of variables to obtain a new representation for our code in (\ref{code}), where the first parity is a systematic node in an equivalent representation.
We start with our $(k,k+2)$ MDS storage code of (\ref{code})
\begin{align}
\left[
\begin{smallmatrix}
{\bf I}_{N} & {\bf 0}_{N} & \ldots & {\bf 0}_{N}\\
{\bf 0}_{N} & {\bf I}_{N} & \ldots & {\bf 0}_{N}\\
\vdots&\vdots&\vdots&\vdots\\
{\bf 0}_{N} & {\bf 0}_{N} & \ldots & {\bf I}_{N}\\
{\bf I}_{N} & {\bf I}_{N} & \ldots & {\bf I}_{N}\\
{\bf A}_1&{\bf A}_2&\ldots & {\bf A}_k
\end{smallmatrix}
\right]{\bf f}.\label{C}
\end{align}
and make the following change of variables 
\begin{align}
\sum_{i=1}^k{\bf f}_i&= {\bf y}_1.\label{y1}\\
{\bf f}_s&={\bf y}_s, \; s\in\{2,\ldots,k\}.\label{ys}
\end{align}
We solve (\ref{y1})  and (\ref{ys}) for ${\bf f}_1$ in terms of the ${\bf y}_i$ variables and obtain
\begin{equation}
{\bf f}_1 = {\bf y}_1-\sum_{s=2}^k{\bf y}_s.\label{f1}
\end{equation}
Then, we plug  (\ref{ys}) and (\ref{f1}) in (\ref{C}), to have the equivalent representation
\begin{align}
\left[
\begin{smallmatrix}
{\bf I}_{N} & -{\bf I}_{N} & \ldots & -{\bf I}_{N}\\
{\bf 0}_{N} & {\bf I}_{N} & \ldots & {\bf 0}_{N}\\
\vdots&\vdots&\vdots&\vdots\\
{\bf 0}_{N} & {\bf 0}_{N} & \ldots & {\bf I}_{N}\\
{\bf I}_{N} & {\bf 0}_{N} & \ldots & {\bf 0}_{N}\\
{\bf A}_1&{\bf A}_2-{\bf A}_1&\ldots & {\bf A}_k-{\bf A}_1.
\end{smallmatrix}
\right]{\bf y}, \label{G}
\end{align}
where ${\bf y} = \left[{\bf y}^T_1\ldots{\bf y}^T_k\right]^T\in\mathbb{F}_q^{kN}$. 
The first parity node of the code in (\ref{code}) now corresponds to the node which contains ${\bf y}_1$ in the aforementioned representation. 
The coding matrices under this new representation are
\begin{align}
{\bf A}_1& = a_1{\bf X}_1+b_1{\bf X}_{k+1}+{\bf I}_{N},\\
{\bf A}_s-{\bf A}_1 & = a_s{\bf X}_s+(b_s-b_1){\bf X}_{k+1}-a_1{\bf X}_1,
\end{align}
for $ s\in\{2,\ldots,k\}$.
In contrast to the systematic node repair process, in the following we use a repair matrix of a slightly different structure.
We construct the repair matrix ${\bf V}_a$ with columns in the set
\begin{equation}
\mathcal{V}_{a} = \left\{\prod_{s=2}^{k+1} \left({\bf X}_1{\bf X}_s\right)^{x_s}{\bf w}:x_s\in\{0,1\}\right\}.
\end{equation}
Observe that this set is also a subset of $\mathcal{H}_{N}$.
Then, to repair the node of (\ref{G}) that contains ${\bf y}_1$ (i.e., the one that corresponds to the first parity node of (\ref{C})) 
we download ${\bf X}_1{\bf V}_a$ times the contents of the first parity in (\ref{G})  and ${\bf V}_a$ times the contents of the second parity.
Hence, during this repair, the useful space is spanned by
\begin{equation}
\left[{\bf X}_1{\bf V}_a \;\; {\bf A}_1{\bf V}_a\right]
\end{equation}
and the interference space due to file part ${\bf y}_s$, $s\in\{2,\ldots,k\}$, is 
\begin{equation}
\left[{\bf X}_1{\bf V}_a \;\; ({\bf A}_s-{\bf A}_1){\bf V}_a\right].
\end{equation}
Before we proceed, observe that the following hold
{\small
\begin{align}
&\mathcal{L}({\bf X}_1{\bf X}_s{\bf V}_a)=\mathcal{L}({\bf V}_a)=\left\{\left(\sum_{s=2}^{k+1}x_s\;(\text{mod }2)\right){\bf e}_1+\sum_{s=2}^{k+1}x_s{\bf e}_s;\; x_s\in\{0,1\}\right\}\\
\Leftrightarrow&\mathcal{L}({\bf X}_1{\bf V}_a)=\mathcal{L}({\bf X}_s{\bf V}_a) =\left\{\left(1+\sum_{s=2}^{k+1}x_s\;(\text{mod }2)\right){\bf e}_1+\sum_{s=2}^{k+1}x_s{\bf e}_s;\; x_s\in\{0,1\}\right\}\\
\Rightarrow&\mathcal{L}({\bf X}_{s_1}{\bf V}_a)=\mathcal{L}({\bf X}_{s_2}{\bf V}_a),
\end{align}
}for any $s,s_1,s_2\in\{1,\ldots,k+1\}$.
The above equations imply that 
\begin{align}
\mathcal{L}\left({\bf V}_a\right)\cup\mathcal{L}\left({\bf X}_1{\bf V}_a\right) &= \left\{\sum_{s=1}^{k+1}x_s;\; x_s\in\{0,1\}\right\}= \mathcal{L}\left({\bf H}_{N}\right).
\end{align}
Therefore, we have the following for each of the interference spaces
{\small
\begin{align}
\frac{N}{2}&\le\rank\left(\left[{\bf X}_1{\bf V}_a\;\; \left(a_s{\bf X}_s+(b_s-b_1){\bf X}_{k+1}-a_1{\bf X}_1\right){\bf V}_a\right]\right)\nonumber\\
&\le\left|\mathcal{L}\left({\bf X}_1{\bf V}_a\right)\cup \mathcal{L}\left({\bf X}_s{\bf V}_a\right) \cup \mathcal{L}\left({\bf X}_{k+1}{\bf V}_a\right)\right|\nonumber\\
&=\left|\mathcal{L}\left({\bf X}_1{\bf V}_a\right)\right|  =\frac{N}{2}.
\end{align}
}Moreover, for the useful data space we have
\begin{align}
\rank\left(\left[{\bf X}_1{\bf V}_a\;\; \left(a_1{\bf X}_1+b_1{\bf X}_{k+1}+{\bf I}_{N}\right){\bf V}_a\right]\right)&=\rank(\left[{\bf X}_1{\bf V}_a \;\; {\bf V}_a\right])\nonumber\\ &=\left|\mathcal{L}\left({\bf V}_a\right)\cup\mathcal{L}\left({\bf X}_1{\bf V}_a\right)\right|=\left|\mathcal{L}\left({\bf H}_{N}\right)\right|= N.
\end{align}
Thus, we can perform optimal repair of the node containing ${\bf y}_1$ in (\ref{G}), which is equivalent to optimally repairing the first parity of our code in (\ref{code}).

\subsection{Repairing the second parity}
Here, we have an additional step. 
We first manipulate our coding matrices of (\ref{code}) to obtain an equivalent representation for the same code. 
Then, in the same manner we rewrite this code in a form where the second parity of (\ref{code})  is a systematic node in some representation.
Without loss of generality, we can multiply any coding column block that multiplies the $i$th file part
\begin{equation}
\left[
\begin{array}{c}
{\bf I}\\
{\bf A}_i
\end{array}
\right]=\left[
\begin{array}{c}
{\bf I}\\
a_i{\bf X}_i+b_i{\bf X}_{k+1}+{\bf I}_{N}
\end{array}
\right]
\end{equation}
with a full rank matrix and maintain the same code properties, as shown in \cite{RashmiProduct}.
In the following derivations, we use the fact that ${\bf X}_s^2 = {\bf I}_{N}$, for any $s\in\{1,\ldots,k+1\}$.
We multiply  the $i$-th block of (\ref{code}) with $a_i{\bf X}_i-b_i{\bf X}_{k+1}+{\bf I}_{N}$ to obtain
{\small
\begin{align}
\left[
\begin{array}{c}
{\bf I}_{N}\\
a_i{\bf X}_i+b_i{\bf X}_{k+1}+{\bf I}_{N}
\end{array} 
\right]\nonumber
&\equiv 
\left[
\begin{array}{c}	
a_i{\bf X}_i-b_i{\bf X}_{k+1}+{\bf I}_{N}\\
\left(a_i{\bf X}_i-b_i{\bf X}_{k+1}+{\bf I}_{N}\right)\left(a_i{\bf X}_i+b_i{\bf X}_{k+1}+{\bf I}_{N}\right)
\end{array}
\right]=
\left[
\begin{array}{c}	
a_i{\bf X}_i-b_i{\bf X}_{k+1}+{\bf I}_{N}\\
\left(a_i{\bf X}_i+{\bf I}_{N}\right)^2-b_i^2{\bf I}_{N}
\end{array}
\right]\nonumber\\
&\equiv
\left[
\begin{array}{c}	
a_i{\bf X}_i-b_i{\bf X}_{k+1}+{\bf I}_{N}\\
2a_i{\bf X}_i+(a_i^2-b_i^2+1){\bf I}_{N}.
\end{array}
\right]\overset{(*)}{=}\left[
\begin{array}{c}	
a_i{\bf X}_i-b_i{\bf X}_{k+1}+{\bf I}_{N}\\
2a_i{\bf X}_i
\end{array}
\right],
\end{align}
}where in $(*)$ we use the fact that $a_i^2-b_i^2+1=0$.
We continue by multiplying the $i$-th column block with $(a_i)^{-1}{\bf X}_i$ to obtain
\begin{align}
\left[
\begin{array}{c}
a_i{\bf X}_i-b_i{\bf X}_{k+1}+{\bf I}_{N}\\
2a_i{\bf X}_i
\end{array} 
\right]
&\equiv
\left[
\begin{array}{c}
{\bf I}_{N}-a_i^{-1}b_i{\bf X}_{k+1}{\bf X}_i+a_i^{-1}{\bf X}_i \\
2{\bf I}_{N}
\end{array}
\right]\equiv
\left[
\begin{array}{c}
{\bf I}_{N}-a_i^{-1}b_i{\bf X}_{k+1}{\bf X}_i+a_i^{-1}{\bf X}_i \\
{\bf I}_{N}
\end{array}
\right],
\end{align}
where in the last step we multiplied the contents of the second parity with $2^{-1}$.
Hence, let
{\small
\begin{align}
{\bf A}'_i&={\bf I}_{N}-a_i^{-1}b_i{\bf X}_{k+1}{\bf X}_i+a_i^{-1}{\bf X}_i, \;i\in\{1,\ldots,k\}.
\end{align}
}Then, we rewrite our original code as
\begin{align}
\left[
\begin{smallmatrix}
{\bf I}_{N} & {\bf 0}_{N} & \ldots & {\bf 0}_{N}\\
{\bf 0}_{N} & {\bf I}_{N} & \ldots & {\bf 0}_{N}\\
\vdots&\vdots&\vdots&\vdots\\
{\bf 0}_{N} & {\bf 0}_{N} & \ldots & {\bf I}_{N}\\
{\bf A}'_1&{\bf A}'_2&\ldots & {\bf A}'_k\\
{\bf I}_{N} & {\bf I}_{N} & \ldots & {\bf I}_{N}
\end{smallmatrix}
\right]{\bf f}'
\end{align}
where ${\bf f}'$ is a full rank row transformation of ${\bf f}$.
We proceed in the same manner that we handled the first parity repair.
We make a change of variables such that the second parity becomes a systematic node in a new representation
\begin{equation}
\sum_{i=1}^k{\bf f}'_i= {\bf y}'_1
\end{equation}
and obtain the equivalent form
\begin{align}
\left[
\begin{smallmatrix}
{\bf I}_{N} & -{\bf I}_{N} & \ldots & -{\bf I}_{N}\\
{\bf 0}_{N} & {\bf I}_{N} & \ldots & {\bf 0}_{N}\\
\vdots&\vdots&\vdots&\vdots\\
{\bf 0}_{N} & {\bf 0}_{N} & \ldots & {\bf I}_{N}\\
{\bf A}'_1&{\bf A}'_2-{\bf A}'_1&\ldots & {\bf A}'_k-{\bf A}'_1\\
{\bf I}_{N} & {\bf 0}_{N} & \ldots & {\bf 0}_{N}.
\end{smallmatrix}
\right]{\bf y}',
\end{align}
where
{\small
\begin{align}
{\bf A}'_1&={\bf I}_{N}-a_1^{-1}b_1{\bf X}_{k+1}{\bf X}_1+a_1^{-1}{\bf X}_1,\\
{\bf A}'_s-{\bf A}'_1&=a_s^{-1}{\bf X}_s-a_s^{-1}b_s{\bf X}_{k+1}{\bf X}_s+a_1^{-1}b_1{\bf X}_{k+1}{\bf X}_1-a_1^{-1}{\bf X}_1
\end{align}
}Then, the parity node which corresponds to systematic node $1$ here, can be repaired by using ${\bf V}_b$ with columns in the set
\begin{equation}
\mathcal{V}_b = \left\{{\bf X}_{k+1}^{x_{k+1}}\prod_{s=2}^{k} \left({\bf X}_1{\bf X}_s\right)^{x_s}{\bf w}:x_{k+1},x_s\in\{0,1\}\right\}.
\end{equation}
Again, the following equations hold
{\small
\begin{align}
&\mathcal{L}({\bf X}_{k+1}{\bf V}_b)=\mathcal{L}({\bf V}_b) = \left\{\left(\sum_{s=2}^kx_s\;(\text{mod }2)\right){\bf e}_1+\sum_{s=2}^{k+1}x_s{\bf e}_s;\; x_s\in\{0,1\}\right\},\\
&\mathcal{L}({\bf X}_{s_1}{\bf V}_b)=\mathcal{L}({\bf X}_{s_2}{\bf V}_b)= \left\{\left(1+\sum_{s=2}^kx_s\;(\text{mod }2)\right){\bf e}_1+\sum_{s=2}^{k+1}x_s{\bf e}_s;\; x_s\in\{0,1\}\right\},\\
\text{and }&\mathcal{L}({\bf X}_{s_1}{\bf X}_{k+1}{\bf V}_b)=\mathcal{L}({\bf X}_{s_1}{\bf V}_b),
\end{align}
}for all $s_1,s_2\in\{1,\ldots,k\}$.
Hence, we have for the interfence space generated by component ${\bf y}_s'$, $s\in\{2,\ldots,k\}$
{\small
\begin{align}
\frac{N}{2}&\le\rank\left(\left[{\bf X_1}{\bf V}_b \;\; ({\bf A}_s'-{\bf A}_1'){\bf V}_b\right)\right)\nonumber\\
&\le|\mathcal{L}\left({\bf X}_s{\bf V}_b\right)\cup \mathcal{L}\left({\bf X}_1{\bf V}_b\right)\cup\mathcal{L}\left({\bf X}_{k+1}{\bf X}_1{\bf V}_b\right)\cup \mathcal{L}\left({\bf X}_{k+1}{\bf X}_s{\bf V}_b\right)|\nonumber\\
&=\left|\mathcal{L}\left({\bf X}_1{\bf V}_b\right)\cup \mathcal{L}\left({\bf X}_s{\bf V}_b\right)\right|=\frac{N}{2}.
\end{align}
Moreover, the useful space is full rank
\begin{align}
\rank\left(\left[{\bf X}_1{\bf V}_b \;\; \left({\bf I}_{N}-a_1^{-1}b_1{\bf X}_{k+1}{\bf X}_1+a_1^{-1}{\bf X}_1\right){\bf V}_b\right]\right)=\rank\left(\left[{\bf X}_1{\bf V}_b \;\; {\bf V}_b\right]\right)=N.
\end{align}
}Thus, we can perform optimal repair for the second parity of the code in (\ref{code}), with repair bandwidth $(k+1)\frac{N}{2}$.

\section{The MDS Property}
In this section, we give explicit conditions on the $a_i, b_i$ constants, for all $i\in\{1,\ldots,k\}$, and the size of the finite field $\mathbb{F}_q$, for which the code in (\ref{code}) is MDS.
We discuss the MDS property using the notion of data collectors (DCs), in the same manner that it was used in \cite{DimakisGWWR:08}.
A DC can be considered as an external user that can connect and has complete access to the contents of some subset of $k$ nodes.
A storage code where each node expends $\frac{M}{k}$ worth of storage, has the MDS property when all possible ${n}\choose{k}$ DCs can decode the file ${\bf f}$.
We can show  that testing the MDS property is equivalent to checking the rank of a specific matrix associated with each DC.
This DC matrix is the vertical concatenation of the $k$ stacks of equations stored by the nodes that the DC connects to.
If all ${n}\choose{k}$ DC matrices are full rank, then we declare that the storage code has the MDS property.

We start with a DC that connects to systematic nodes $\{1,\ldots,k-1\}$ and the first parity node.
The determinant of the corresponding DC matrix is
{\small
\begin{equation}
\begin{split}
\det\left(\left[
\begin{array}{ccc|c}
{\bf I}_{N} &\ldots & {\bf 0}_{N\times N}&{\bf 0}_{N\times N}\\
\vdots&&\vdots&\vdots\\
{\bf 0}_{N\times N} &\ldots &  {\bf I}_{N}&{\bf 0}_{N\times N}\\
\hline
{\bf I}_{N} &\ldots&{\bf I}_{N}&{\bf I}_{N}\\
\end{array}
\right]\right)
=\det\left({\bf I}_{N}\right)\ne0,
\end{split}
\end{equation}
}since ${\bf I}_{N}$ is a full rank diagonal matrix. 
We continue by considering a DC that connects to systematic nodes $\{1,\ldots,k-1\}$ and the second parity node.
For that we have
{\small
\begin{equation}
\begin{split}
&\det\left(\left[
\begin{array}{ccc|c}
{\bf I}_{N} &\ldots & {\bf 0}_{N\times N}&{\bf 0}_{N\times N}\\
\vdots&&\vdots&\vdots\\
{\bf 0}_{N\times N} &\ldots &  {\bf I}_{N}&{\bf 0}_{N\times N}\\
\hline
{\bf A}_1 &\ldots&{\bf A}_{k-1}&{\bf A}_k\\
\end{array}
\right]\right)=\det\left({\bf A}_k\right)\ne0,
\end{split}
\end{equation}
}due to ${\bf A}_k$ being full rank.

Finally, we consider DCs that connect to $k$ systematic nodes and both parity nodes.
Let a DC that connects to systematic node $\{1,\ldots,k-2\}$ and the two parities.
The corresponding DC matrix is
{\small
\begin{equation}
\left[
\begin{array}{ccc|cc}
{\bf I}_{N} & \ldots & {\bf 0}_{N\times N}&{\bf 0}_{N\times N}&{\bf 0}_{N\times N}\\
\vdots&&\vdots&\vdots\\
{\bf 0}_{N\times N} &\ldots &  {\bf I}_{N}&{\bf 0}_{N\times N}&{\bf 0}_{N\times N}\\
\hline
{\bf I}_{N} &\ldots&{\bf I}_{N}&{\bf I}_{N}&{\bf I}_{N}\\
{\bf A}_1 &\ldots&{\bf A}_{k-2}&{\bf A}_{k-1}&{\bf A}_k
\end{array}
\right].
\label{DC_2}
\end{equation}
}The leftmost $(k-2)N$ columns of the matrix in (\ref{DC_2}) are linearly independent, due to the upper-left identity block.
Moreover, the leftmost $(k-2)N$ columns are linearly independent with the rightmost $2N$, using an analogous argument.
Hence, we need to only check the rank of the sub-matrix
\begin{equation}
\left[
\begin{array}{cc}
{\bf I}_{N}&{\bf I}_{N}\\
{\bf A}_{k-1}&{\bf A}_k
\end{array}
\right].
\end{equation}
In the general case, a DC that connects to some $k-2$ subset of systematic nodes and the two parities has a corresponding matrix where the following block needs to be full rank so that the MDS property can be satisfied
\begin{equation}
\left[
\begin{array}{cc}
{\bf I}_{N}&{\bf I}_{N}\\
{\bf A}_{i}&{\bf A}_j
\end{array}
\right],
\end{equation}
for $i,j\in\{1,\ldots,k\}$ and $i\ne j$.
The code is MDS when 
{\small
\begin{align}
&\rank\left(\left[
\begin{array}{cc}
{\bf I}_{N}&{\bf I}_{N}\\
a_i{\bf X}_i+b_i{\bf X}_{k+1}+{\bf I}_{N}&a_j{\bf X}_j+b_j{\bf X}_{k+1}+{\bf I}_{N}
\end{array}
\right]\right)\nonumber\\
&=\rank\Biggl(\left[
\begin{array}{cc}
{\bf I}_{N}&{\bf I}_{N}\\
a_i{\bf X}_i+b_i{\bf X}_{k+1}+{\bf I}_{N}&a_j{\bf X}_j+b_j{\bf X}_{k+1}+{\bf I}_{N}
\end{array}
\right]\times \left[
\begin{array}{cc}
{\bf I}_{N}&{\bf I}_{N}\\
{\bf 0}_{N\times N}&-{\bf I}_{N}
\end{array}
\right]
\Biggr)\nonumber\\
&=\rank\left(\left[
\begin{smallmatrix}
{\bf I}_{N}&0\\
a_i{\bf X}_i+b_i{\bf X}_{k+1}+{\bf I}_{N}& a_i{\bf X}_i-a_j{\bf X}_j+(b_i-b_j){\bf X}_{k+1}
\end{smallmatrix}
\right]\right)\nonumber\\
&=\frac{N}{2}+\rank\left(a_i{\bf X}_i-a_j{\bf X}_j+(b_i-b_j){\bf X}_{k+1}\right)=N,
\end{align}
}for all $i,j\in\{1,\ldots,k\}$, which is true if 
\begin{equation}
\rank\left(a_i{\bf X}_i-a_j{\bf X}_j+(b_i-b_j){\bf X}_{k+1}\right)=\frac{N}{2}.
\end{equation}
Since the diagonal elements of ${\bf X}_i$ are $\{\pm1\}$, the previous requirement gives the lemma.
\begin{lem}
The code in (\ref{code}) is MDS when
\begin{align}
i)&\;\;a_i- a_j + (b_i-b_j)\ne0,\label{mds1}\\
ii)&\;\;a_i+ a_j - (b_i-b_j)\ne0,\\
iii)&\;\;a_i- a_j - (b_i-b_j)\ne0,\\
\text{ and }iv)&\;\;a_i+ a_j + (b_i-b_j)\ne0,\label{mds4}
\end{align}
for all $i\ne j\in\{1,\ldots, k\}$.
\end{lem}
Now, remember that our initial constraint on the $a_i$ and $b_i$ constants was 
\begin{equation}
a_i^2-b_i^2 = -1\Leftrightarrow (a_i-b_i)(a_i+b_i)= -1. \label{sqrdiff}
\end{equation}
one solution to the previous equation is the following
\begin{align}
a_i-b_i &= x_i\\
a_i+b_i &=-x_i^{-1},
\end{align}
If we input the above solution to (\ref{sqrdiff}), then the MDS equations (\ref{mds1})-(\ref{mds4}) become
\begin{align}
 a_i- a_j + (b_i-b_j) &=  a_i+ b_i - (a_i+b_j)\nonumber\\
&= -x_i^{-1}+x_j^{-1}\ne 0\nonumber\\
\Leftrightarrow &x_i^{-1}\ne x_j^{-1},\\
a_i+ a_j - (b_i-b_j) &=  a_i- b_i + a_j+b_j\nonumber\\
& = x_i-x_j^{-1}\ne 0\nonumber\\
\Leftrightarrow &x_i\ne x_j^{-1},\\
a_i- a_j - (b_i-b_j)& =  a_i- b_i - (a_j-b_j)\nonumber\\
& = x_i-x_j\ne 0\nonumber\\
\Leftrightarrow &x_i\ne x_j,\\
a_i+ a_j + (b_i-b_j) &=  a_i+ b_i +a_j-b_j\nonumber\\
& = -x_i^{-1}+x_j\ne 0\nonumber\\
\Leftrightarrow &x_i^{-1}\ne x_j,
\end{align}
The above conditions can be equivalently stated as
\begin{equation}
x_i\ne x_j \text{ and } x_ix_j\ne 1,
\end{equation}
for any $i\ne j \in\{1,\ldots,k\}$.

Then, consider a prime field $\mathbb{F}_q$ of size $q$. 
The set of $x_i$s that satisfies our MDS requirements, is such in which no two elements are inverses of each other.
It is known that, over a prime field,  half the nonzero elements are inverses of the other nonzero half.
If we additionally do not consider $x_i\in\{1,q-1\}$, then we are left with $\frac{q-3}{2}$ elements.
Therefore, we can consider a prime field of size $q$ that has the property
\begin{equation}
k\le\frac{q-3}{2} \Leftrightarrow q\ge 2k+3
\end{equation}
and obtain $x_1,\ldots,x_k$ such that our requirements are satisfied. 
Then, the elements $a_i$ and $b_i$, for all $i\in\{1,\ldots,k\}$, can be obtained through the following equations
\begin{align}
a_i& = 2^{-1}x_i-2^{-1}x_i^{-1}\\
b_i& = -2^{-1}x_i-2^{-1}x_i^{-1}.
\end{align}
Observe that the above solutions yield $a_i\ne 0$ (that is needed for successful repair), for all $i\in\{1,\ldots,k\}$, when $x_i\notin\{0,1,q-1\}$. 
Therefore a prime field of size greater than, or equal to $2k+3$ always suffices to obtain the MDS property.

\section{Generalizing to more than $2$ parities}
\subsection{$m$-parity codes with optimal systematic repair}

We generalize the Hadamard design construction of Section III and of the code in \cite{PD1}, to construct $(k+m,k)$ MDS storage codes for file sizes $M= km^k$.
Our constructions are based on a generalization of the Sylvester construction for complex Hadamard matrices that use $m^{\text{th}}$ roots of unity.
We generate these matrices as
\begin{equation}
{\bf H}_{m^k} = {\bf H}_{m}\otimes{\bf H}_{m^{k-1}},
\end{equation}
where ${\bf H}_m$ is the $m$-point Discrete Fourier Transform matrix over a finite field. 
For example, for $m=3$ and $\mathbb{F}_{7}$, we have
\begin{equation}
{\bf H}_3 = \left[
\begin{smallmatrix}
1 & 1& 1\\
1 & \rho& \rho^2\\
1 & \rho^2& \rho
\end{smallmatrix}
\right] \text{ and } {\bf H}_9 = \left[
\begin{smallmatrix}
{\bf H}_3 & {\bf H}_3 & {\bf H}_3\\
{\bf H}_3 & \rho{\bf H}_3& \rho^2{\bf H}_3\\
{\bf H}_3 & \rho^2{\bf H}_3& \rho{\bf H}_3
\end{smallmatrix}
\right],
\end{equation}
where $\rho = 2$.
Then, we consider the set
\begin{equation}
\mathcal{H}_{m^k} = \left\{\prod_{i = 1}^{k}{\bf X}_{i}^{x_i}{\bf w}: x_{i}\in \{0,1,\ldots, m-1\}\right\}, \label{Hprodm}
\end{equation}
where ${\bf w}={\bf 1}_{m^k\times 1}$ and
{\small
\begin{equation}
{\bf X}_i = {\bf I}_{m^{i-1}}\otimes \text{blkdiag}\left({\bf I}_{\frac{N}{m^{i}}},\rho{\bf I}_{\frac{N}{m^{i}}},\ldots,\rho^{m-1}{\bf I}_{\frac{N}{m^{i}}}\right).
\end{equation}
}Here, $\rho$ denotes an $m^{\text{th}}$ root of unity which yields
\begin{equation}
{\bf X}^m_i = {\bf I}_{m^k}.
\end{equation}
As with the $m=2$ case, there is a one-to-one correspondence between the elements of the set $\mathcal{H}_{m^k}$ and the columns of ${\bf H}_{m^k}$.
The general $m$ proof for that property follows the same manner of the $m=2$ case, thus we omit it.
\begin{rem}
To maintain the full rank property of ${\bf H}_{m^k}$, the finite field over which we operate should be chosen such that all $m^{\text{th}}$ roots of unity are distinct.
The number of distinct $m^{\text{th}}$ roots of unity over a finite field $\mathbb{F}_q$ is given by the number of (distinct) solutions of the equation $x^m = 1$.
This is equal to the order of the cyclic group that generates $m^{\text{th}}$ roots of unity within the multiplicative group of $\mathbb{F}_q$. This subgroup has order $m$ when $m$ divides $q-1$ \cite{Fields}.
\end{rem}

\subsubsection{Code construction}
Our $(k+m,k)$ MDS code encodes a file ${\bf f}$ of size $M = km^k$ in the manner of
\begin{equation}
\left[
\begin{array}{c}
{\bf I}_{km^k}\\
{\bf A}^{(k,m)}
\end{array}
\right]{\bf f},\label{mcode}
\end{equation}
where 
\begin{equation}
{\bf A}^{(k,m)} = 
\left[
\begin{array}{cccc}
{\bf I}_{m^k}&{\bf I}_{m^k}&\ldots&{\bf I}_{m^k}\\
\lambda_{1,1}{\bf X}_1 & \lambda_{1,2}{\bf X}_2 & \ldots & \lambda_{1,k}{\bf X}_k\\
\lambda_{2,1}{\bf X}^2_1 & \lambda_{2,2}{\bf X}^2_2 & \ldots & \lambda_{2,k}{\bf X}^2_k\\
\vdots\\
\lambda_{m-1,k}{\bf X}^{m-1}_1 & \lambda_{m-1,2}{\bf X}^{m-1}_2 & \ldots & \lambda_{m-1,k}{\bf X}^{m-1}_k
\end{array}
\right],
\end{equation}
with $\lambda_{i,j}\in\mathbb{F}_q$.

\subsubsection{Optimal repair of the systematic nodes}
For this code, let systematic node $i\in\{1,\ldots,k\}$ fail.
Then, to repair it we construct the repair matrix ${\bf V}_i$ that has as columns the elements of set
\begin{equation}
\mathcal{V}_i = \left\{\prod_{s = 1,s\ne i}^{k}{\bf X}_{s}^{x_s}{\bf w}: x_{s}\in \{0,1,\ldots, m-1\}\right\}.
\end{equation}
This matrix is used to multiply the contents of each of the parity nodes.
Here, the useful space during the repair is given by 
\begin{equation}
\left[{\bf V}_i\;\;\;{\bf X}_i{\bf V}_i\;\;\;{\bf X}^2_i{\bf V}_i\;\ldots\;{\bf X}^{m-1}_i{\bf V}_i\right]
\end{equation}
and the interference space generated by systematic component $j\ne i$ is spanned by
\begin{equation}
\left[{\bf V}_i\;\;\;{\bf X}_j{\bf V}_i\;\;\;{\bf X}^2_j{\bf V}_i\;\ldots\;{\bf X}^{m-1}_j{\bf V}_i\right].
\end{equation}
Due to the modulus-$m$ property of the powers of the ${\bf X}_i$ matrices, we obtain the following under the lattice representation
\begin{equation}
\mathcal{L}\left({\bf V}_i\right)=\mathcal{L}\left({\bf X}^l_j{\bf V}_i\right) \text{ and }\mathcal{L}\left({\bf X}^{l_1}_{i}{\bf V}_i\right)\cap\mathcal{L}\left({\bf X}^{l_2}_i{\bf V}_i\right)=\emptyset,
\end{equation}
for any $j\in\{1,\ldots k\}\ne i$, and $l,l_1,l_2\in \{0,\ldots,m-1\}$, with $l_1\ne l_2$.
The above property and the fact that the elements of $\mathcal{H}_{m^k}$ are linearly independent leads us to the following lemma.
\begin{lem}
For any $i,j\in\{1,2,\ldots,k\}$ we have that
\begin{align}
\rank(\left[{\bf V}_i\;\;\;{\bf X}_j{\bf V}_i\;\;\;{\bf X}^2_j{\bf V}_i\;\ldots\;{\bf X}^{m-1}_j{\bf V}_i\right])&=\left|\mathcal{L}({\bf V}_i)\cup\mathcal{L}\left({\bf X}_j{\bf V}_i\right)\cup\mathcal{L}\left({\bf X}^{2}_j{\bf V}_i\right)\cup\ldots \cup\mathcal{L}\left({\bf X}^{m-1}_j{\bf V}_i\right)\right|\\
&=\left\{
\begin{array}{lc}
m^k, & i=j,\\
m^{k-1}, & i\ne j.
\end{array}
\right.
\end{align}
\label{mlattice}
\end{lem}
By Lemma (\ref{mlattice}) we see that each of the $k-1$ interference terms is confined within $m^{k-1}$ dimensions and the full rank property of the useful space is maintained.
This is equivalent to stating that we can repair a single systematic node failure by downloading exactly $m^k+(k-1)m^{k-1}=(n-1)m^{k-1}$ equations, which matches exactly the information theoretic repair optimal of \cite{DimakisGWWR:08}.

In Fig. 9 we give an illustration of the repair spaces for a $(6,3)$ code.
We sketch the structure of our code on the left of the figure. 
Each parity block is associated with a specific {\it key} matrix ${\bf X}_i$. 
This allows a selection of ${\bf V}_i$ that is an invariant subspace to all matrices but to the key, one which multiplies the desired and lost file piece.
This selection of ${\bf V}_i$ results in perfect alignment of interference in $3^2$ dimensions, while ensuring a full rank  $3^3$ useful space.

\begin{figure}[h]
\centerline{\includegraphics[width=0.32\columnwidth]{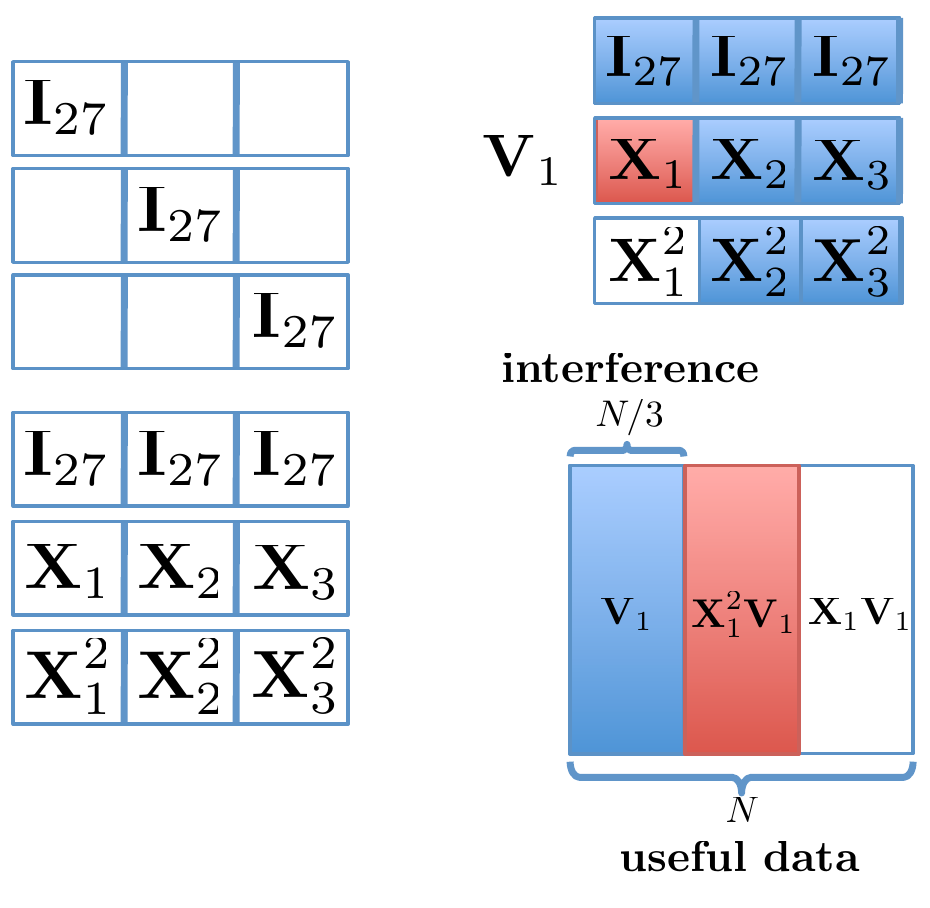}}
\caption{A $(6,3)$ systematic-repair optimal code.}
\label{code63}
\end{figure}

\subsubsection{Suboptimal repair of the parities}
In contrast to our $2$-parity code of (\ref{code}), for this $m$-parity code, a parity node failure is repaired using the scheme of Wu {\it et al.} \cite{WuD:09}.
We first rewrite our code in a new systematic re-interpretation, where the lost parity is now in systematic form, in the same manner of the parity repair of our $2$-parity code. 
During the repair, we align a single interference block by inverting the corresponding matrices.
This induces a repair download of $m^{k-1}+(n-2)m^k$ equations, which suffices to exactly reconstruct what was lost. 
This repair strategy is only optimal for $(n,2)$ codes and asymptotically matches the file size for large $k$.

\subsubsection{The MDS property}
We establish the MDS property of our $m$-parity codes in a probabilistic sense: we show that when we select the $\lambda_{i,j}$ variables uniformly at random over a sufficiently large finite field, then the code is MDS with probability arbitrarily close to $1$. 
This is shown using the Schwartz-Zippel lemma \cite{Ho, Motwani} on a nonzero polynomial on $\lambda_{i,j}$s induced by the products of all possible DC matrix determinants.

Let a DC of the code in (\ref{mcode}) that connects to $k-p$ systematic nodes  and $p$ parities. 
For simplicity consider that this is the DC that is connected to the last $k-p$ systematic nodes and the first $p$ parity nodes.
The induced determinant of the corresponding DC matrix will be zero if the following determinant is zero
{\small
\begin{align}
&\det\left(\left[
\begin{smallmatrix}
&{\bf 0}_{(k-p)m^k \times pm^k}& {\bf I}_{(k-p)m^k} &\\
\hline
{\bf I}_{m^k}&{\bf I}_{m^k}&\ldots&{\bf I}_{m^k}\\
\lambda_{1,1}{\bf X}_1 & \lambda_{1,2}{\bf X}_2 & \ldots & \lambda_{1,k}{\bf X}_{k}\\
\lambda_{2,1}{\bf X}^2_1 & \lambda_{2,2}{\bf X}^2_2 & \ldots & \lambda_{2,k}{\bf X}^2_{k}\\
\vdots\\
\lambda_{p-1,1}{\bf X}^{p-1}_1 & \lambda_{p-1,2}{\bf X}^{p-1}_2 & \ldots & \lambda_{p-1,k}{\bf X}^{p-1}_{k}
\end{smallmatrix}
\right]\right) = |{\bf I}_{(k-p)m^k}|\det\left(\left[
\begin{smallmatrix}
{\bf I}_{m^k}&{\bf I}_{m^k}&\ldots&{\bf I}_{m^k}\\
\lambda_{1,1}{\bf X}_1 & \lambda_{1,2}{\bf X}_2 & \ldots & \lambda^{(1)}_{p}{\bf X}_{p}\\
\lambda_{2,1}{\bf X}^2_1 & \lambda_{2,2}{\bf X}^2_2 & \ldots & \lambda^{(2)}_{p}{\bf X}^2_{p}\\
\vdots\\
\lambda_{p-1,1}{\bf X}^{p-1}_1 & \lambda_{p-1,2}{\bf X}^{p-1}_2 & \ldots &\lambda_{p-1,p}{\bf X}^{p-1}_{p}
\end{smallmatrix}
\right]\right).\label{mDC}
\end{align}
}Since each of the ${\bf X}_i$ matrices is diagonal, each column of the matrix in the right hand side of (\ref{mDC}) has exactly $p$ nonzero elements. 
These, $pm^k$ columns can be considered to fall into $m^k$ groups,  with each element of a group having identical non-zero support with any other vector in that group. 
Then, any two columns within a block
\begin{equation}
\left[
\begin{smallmatrix}
{\bf I}_{m^k}\\
\lambda_{1,i}{\bf X}_i\\
\lambda_{2,i}{\bf X}^2_i\\
\vdots\\
\lambda_{p-1,i}{\bf X}^{p-1}_i
\end{smallmatrix}
\right]
\end{equation}
are orthogonal since their nonzero supports have zero overlap.
Hence, a linear dependence will only exist among columns of a given non-zero support.
We can then rewrite the matrix determinant of (\ref{mDC}) as
\begin{equation}
\det\left({\bf P}_r\left[
\begin{smallmatrix}
{\bf B}_1&{\bf 0}_{m^k\times m^k}&\ldots&{\bf 0}_{m^k\times m^k}\\
{\bf 0}_{m^k\times m^k}&{\bf B}_2&\ldots&{\bf 0}_{m^k\times m^k}\\
\vdots\\
{\bf 0}_{m^k\times m^k}&\ldots&{\bf 0}_{m^k\times m^k}&{\bf B}_p
\end{smallmatrix}
\right]{\bf P}_c\right)=|{\bf P}_r||{\bf P}_c|\prod_{i=1}^{m^k}|{\bf B}_i|
\end{equation}
where ${\bf P}_r$ and ${\bf P}_c$ are the permutation matrices that group the columns and rows of the matrix according to their non-zero support so to generate the block diagonal matrix. 
The $p\times p$ matrix ${\bf B}_i$ is of the form
\begin{equation}
\left[
\begin{smallmatrix}
\rho_{i_1,j_1}\lambda_{i_1,j_1}& \rho_{i_1,j_2}\lambda_{i_1,j_2} &\ldots &\rho_{i_1,j_p}\lambda_{i_1,j_p} \\
\rho_{i_2,j_1}\lambda_{i_2,j_1}& \rho_{i_2,j_2}\lambda_{i_2,j_2} &\ldots &\rho_{i_2,j_p}\lambda_{i_2,j_p} \\
\vdots & \vdots & \ddots & \vdots \\
\rho_{i_p,j_1}\lambda_{i_p,j_1}& \rho_{i_p,j_2}\lambda_{i_p,j_2} &\ldots &\rho_{i_p,j_p}\lambda_{i_p,j_p}
\end{smallmatrix}
\right].
\end{equation}
where $\rho_{i_1,j_1}$ is some $m^{\text{th}}$ root of unity, the indeces depend on $i$, and no $\lambda_{i,j}$ appears more than once within each matrix.
We can expand the determinant of any ${\bf B}_i$ matrix using the Leibniz formula, where $p!$ monomials of degree $p$ appear.
Each one of them includes a different subset of the $\lambda_{i,j}$ variables. Hence, the induced polynomial cannot be the zero polynomial.
Therefore, the determinant of ${\bf B}_i$ is a nonzero polynomial of degree $p$ in the $\lambda_{i,j}$ variables, hence $\prod_{i=1}^{m^k}|{\bf B}_i|$ is also a non zero polynomial of degree $pm^k$ in the $\lambda_{i,j}$  variables.
Accordingly, we can compute the determinant of each DC in this way. 
In the same manner, each of them will be a nonzero polynomial in $\lambda_{i,j}$. 
The product of all these determinants be a nonzero polynomial in $\lambda_{i,j}$ of some degree $d$.
By the  Schwartz-Zippel lemma \cite{Motwani}, we know that when we draw $\lambda_{i,j}$ uniformly at random over a field of size $q$, this induced polynomial is zero with probability less than or equal to $\frac{d}{q}$.
Hence, the MDS property is satisfied with probability arbitrarily close to $1$, for sufficiently large finite fields.

\section{Connection to Permutation-Matrix Based Codes}
Here we investigate some interesting connections between our systematic-repair optimal codes of Section VIII and the permutation-matrix based codes presented in \cite{Tamo} and \cite{PermCodes}.
Under a similarity transformation, our codes are equivalent to ones with coding matrices picked as specific permutation matrices.
Multiplying the column space of an ${\bf X}_i$ matrix of our construction with the Hadamard matrix ${\bf H}_{m^k}$, yields a matrix that is a permutation of the columns of the Hadamard matrix
\begin{equation}
{\bf H}^{-1}_{m^k}{\bf X}_i {\bf H}_{m^k} = {\bf H}^{-1}_{m^k}{\bf H}_{m^k}{\bf P}_i={\bf P}_i,
\end{equation}
where ${\bf P}_i$ is some permutation matrix.
This is due to the fact that the elements of $\mathcal{H}_{m^k}$ wrap around, i.e.,
 $\mathcal{L}({\bf H}_{m^k})=\mathcal{L}({\bf X}_i{\bf H}_{m^k})$ for any $i$.
\begin{example}
Consider the $m=2,k=3$ case:
\begin{align}
{\bf H}_{2^3} &= \left[{\bf w}\;\;\;{\bf X}_2{\bf w}\;\;\;{\bf X}_1{\bf w}\;\;\;{\bf X}_1{\bf X}_2{\bf w}\right]\\
{\bf X}_1{\bf H}_{2^3}& = \left[{\bf X}_1{\bf w}\;\;\;{\bf X}_1{\bf X}_2{\bf w}\;\;\;{\bf w}\;\;\;{\bf X}_2{\bf w}\right]={\bf H}_{2^3} {\bf P}_1\\
{\bf X}_2{\bf H}_{2^3} &= \left[{\bf X}_2{\bf w}\;\;\;{\bf w}\;\;\;{\bf X}_1{\bf X}_2{\bf w}\;\;\;{\bf X}_1{\bf w}\right]={\bf H}_{2^3} {\bf P}_2,
\end{align}
where ${\bf P}_1$ and ${\bf P}_2$ are permutation matrices. 
The wrap-round property of the columns of the Hadamard matrix produces permutations of itself when multiplied by the ${\bf X}_i$ matrices, and each permutation is distinct.
\end{example}
Without loss of generality \cite{ShahRKR}, we can rewrite the ${\bf A}^{(m,k)}$ matrix of (\ref{mcode}) as
{\small
\begin{align}
{\bf H}^{-1}_{m^k}{\bf A}^{(k,m)}\left({\bf I}_{k}\otimes {\bf H}_{m^k}\right)& = 
{\bf H}^{-1}_{m^k}\left[
\begin{smallmatrix}
{\bf I}_{m^k}{\bf H}_{m^k}&{\bf I}_{m^k}{\bf H}_{m^k}&\ldots&{\bf I}_{m^k}{\bf H}_{m^k}\\
\lambda_{1,1}{\bf X}_1{\bf H}_{m^k} & \lambda_{1,2}{\bf X}_2{\bf H}_{m^k} & \ldots & \lambda_{1,k}{\bf X}_k{\bf H}_{m^k}\\
\lambda_{2,1}{\bf X}^2_1 {\bf H}_{m^k} & \lambda_{2,2}{\bf X}^2_2 {\bf H}_{m^k} & \ldots & \lambda_{2,k}{\bf X}^2_k {\bf H}_{m^k}\\
\vdots\\
\lambda_{m-1,k}{\bf X}^{m-1}_1{\bf H}_{m^k} & \lambda_{m-1,2}{\bf X}^{m-1}_2{\bf H}_{m^k} & \ldots & \lambda_{m-1,k}{\bf X}^{m-1}_k{\bf H}_{m^k}
\end{smallmatrix}
\right]\nonumber\\
&={\bf H}^{-1}_{m^k}\left[
\begin{smallmatrix}
{\bf H}_{m^k}&{\bf H}_{m^k}&\ldots&{\bf H}_{m^k}\\
\lambda_{1,1}{\bf H}_{m^k}{\bf P}_{1,1} & \lambda_{1,2}{\bf H}_{m^k}{\bf P}_{1,2} & \ldots & \lambda_{1,k}{\bf H}_{m^k}{\bf P}_{1,k}\\
\lambda_{2,1}{\bf H}_{m^k}{\bf P}_{2,1} & \lambda_{2,2}{\bf H}_{m^k}{\bf P}_{2,2} & \ldots & \lambda_{2,k}{\bf H}_{m^k}{\bf P}_{2,k}\\
\vdots\\
\lambda_{m-1,k}{\bf H}_{m^k}{\bf P}_{m-1,1} & \lambda_{m-1,2}{\bf H}_{m^k}{\bf P}_{m-1,2} & \ldots & \lambda_{m-1,k}{\bf H}_{m^k}{\bf P}_{m-1,k}
\end{smallmatrix}
\right]\nonumber\\
&=\left[
\begin{smallmatrix}
{\bf I}_{m^k}&{\bf I}_{m^k}&\ldots&{\bf I}_{m^k}\\
\lambda_{1,1}{\bf P}_{1,1} & \lambda_{1,2}{\bf P}_{1,2} & \ldots & \lambda_{1,k}{\bf P}_{1,k}\\
\lambda_{2,1}{\bf P}_{2,1} & \lambda_{2,2}{\bf P}_{2,2} & \ldots & \lambda_{2,k}{\bf P}_{2,k}\\
\vdots\\
\lambda_{m-1,k}{\bf P}_{m-1,1} & \lambda_{m-1,2}{\bf P}_{m-1,2} & \ldots & \lambda_{m-1,k}{\bf P}_{m-1,k}
\end{smallmatrix}
\right],
\end{align}
}where ${\bf P}_{i,j}$ is a permutation matrix.
The systematic nodes of this equivalent $(k+m,m)$ MDS code can be optimally repaired using the repair matrices ${\bf V}_i{\bf H}_i^{-1}$, where ${\bf V}_i$ has the columns of the set
$\mathcal{V}_i = \left\{\prod_{s = 1,s\ne i}^{k}{\bf X}_{s}^{x_s}{\bf w}: x_{s}\in \{0,1,\ldots, m-1\}\right\}$.
This is true since the rank properties of the correspoding useful and interference spaces remain the same under full rank column transformations.
Interestingly, this connection is two-way.
We give an example of a permutation code from \cite{PermCodes} that exactly maps to our designs. 
\begin{example}
We consider the $(5,3)$ permutation code of \cite{PermCodes}, designed for file sizes $M = 3\cdot 2^3$.
The three coding matrices of the first parity of this code are three identity matrices ${\bf I}_8$.
The three coding matrices of the second parity  are three permutation matrices
\begin{align}
{\bf P}_1 &= {\bf I}_{\{5, 6, 7, 8, 1, 2, 3, 4\},:},\;\;{\bf P}_2 = {\bf I}_{\{3,4,1,2,7,8,5,6\},:},\text{ and}\;\;{\bf P}_3  = {\bf I}_{\{2,1,4,3,6,5,8,7\},:},
\end{align}
where ${\bf I}_{\{i_1, i_2, i_3, i_4, i_5, i_6, i_7, i_18\},:}$ indicates a permutation of the columns of the $8\times 8$ identity matrix.
We know that these matrices commute, therefore since they are normal, they can be simultaneously diagonalized under a common eigen basis.
It can be checked that a common basis for the above commuting permutation matices is the Hadamard matrix, which gives
\begin{align}
{\bf H}_8{\bf P}_1{\bf H}_8^T={\bf X}_1,\;{\bf H}_8{\bf P}_2{\bf H}_8^T={\bf X}_2,\;{\bf H}_8{\bf P}_3{\bf H}_8^T={\bf X}_3.
\end{align}
\end{example}

The connection manifested by the above equivalence examples seems very interesting.
We believe that further investigation on it can lead to better understanding of the repair optimal high-rate MDS code regime.

\section{Conculsions}
We presented the first explicit, high-rate, $(k+2,k)$ erasure MDS storage code that achieves optimal repair bandwidth for any single node failure, including the parities.
Our construction is based on perfect interference alignment properties offered by Hadamard designs.
We generalize our $2$-parity constructions to erasure codes with $m$-parities that achieve optimal repair of the systematic parts.

\section{Acknowledgement}
The authors would like to thank Changho Suh for insightful discussions.

\section*{Appendix} 

{\bf Proof of Lemma 1}: Observe that ${\bf H}_{N} = {\bf H}^T_{N}$ and
\begin{align}
{\bf H}_{N}{\bf H}_{N}^T ={\bf H}_{N}{\bf H}_{N}=\left[
\begin{array}{cc}
2{\bf H}_{\frac{N}{2}}{\bf H}_{\frac{N}{2}} & {\bf 0}_{\frac{N}{2}\times \frac{N}{2}}\\
{\bf 0}_{\frac{N}{2}\times \frac{N}{2}} & 2{\bf H}_{\frac{N}{2}}{\bf H}_{\frac{N}{2}}
\end{array}
\right] &= 2\left({\bf I}_{2}\otimes{\bf H}_{\frac{N}{2}}{\bf H}_{\frac{N}{2}}\right)\nonumber\\
&= 2\left({\bf I}_{2}\otimes2\left({\bf I}_{2}\otimes{\bf H}_{\frac{N}{2}}{\bf H}_{\frac{N}{4}}\right)\right) \nonumber\\
&= 4\left({\bf I}_{4}\otimes{\bf H}_{\frac{N}{4}}{\bf H}_{\frac{N}{4}}\right)\nonumber\\
&\hspace{0.17cm} \vdots\nonumber\\
& = N\cdot\left({\bf I}_{N}\otimes{\bf H}_{1}{\bf H}_{1}\right) = N\cdot{\bf I}_{N}.
\end{align}
We also have that $N \ne 0 \;(\text{mod }q)$, for $q>2$, thus the rank of ${\bf H}_{N}$ is $N$ and its columns are mutually orthogonal.
\hfill$\Box$\\
Then, let an $N\times N$ diagonal matrix
\begin{equation}
{\bf X}_i = {\bf I}_{2^{i-1}}\otimes \text{blkdiag}\left({\bf I}_{\frac{N}{2^{i}}},-{\bf I}_{\frac{N}{2^{i}}}\right)
\end{equation}
defined for $i=\{1,\ldots,\log_2(N)\}$.
${\bf X}_i$ is a diagonal matrix, whose elements is a series of alternating $1$s and $-1$s, starting with $\frac{N}{2^{i}}$ $1$s that flip to $-1$s and back every $\frac{N}{2^{i}}$ positions.
We can now expand ${\bf H}_{N}$ in the following way
\begin{align}
{\bf H}_{N} &= \left[
\begin{array}{cr}
{\bf H}_{\frac{N}{2}}&{\bf H}_{\frac{N}{2}}\\
{\bf H}_{\frac{N}{2}}&-{\bf H}_{\frac{N}{2}}
\end{array}
\right]=\left[\underbrace{{\bf 1}_{2\times1}\otimes{\bf H}_{\frac{N}{2}}}_{{\bf F}_1} \;\;\;{\bf X}_{1}\left({\bf 1}_{2\times1}\otimes{\bf H}_{\frac{N}{2}}\right)\right].
\end{align}
We proceed in the same manner by expanding all ``smaller'' ${\bf H}_{\frac{N}{2^i}}$s
{\small
\begin{align}
{\bf F}_1=&
{\bf 1}_{2\times1}\otimes\left[{\bf 1}_{2\times1}\otimes{\bf H}_{\frac{N}{2^2}} \;\;\;{\bf X}_1\left({\bf 1}_{2\times1}\otimes{\bf H}_{\frac{N}{2^2}}\right)\right]
=\left[\underbrace{{\bf 1}_{2^2\times1}\otimes{\bf H}_{\frac{N}{2^2}}}_{{\bf F}_2} \;\;\;{\bf X}_2\left({\bf 1}_{2^2\times1}\otimes{\bf H}_{\frac{N}{2^2}}\right) \right]\nonumber\\
{\bf F}_2& = \left[\underbrace{{\bf 1}_{2^3\times1}\otimes{\bf H}_{\frac{N}{2^3}}}_{{\bf F}_3} \;\;\;{\bf X}_3\left({\bf 1}_{2^3\times1}\otimes{\bf H}_{\frac{N}{2^3}}\right) \right]\nonumber\\
&\hspace{0.17cm}\vdots\nonumber\\
{\bf F}_{\log_2(N)-1}
&=\left[{\bf 1}_{N\times1} \;\;\;{\bf X}_{\log_2(N)}{\bf 1}_{N\times1} \right],
\end{align}
}where ${\bf F}_i$ is an $N\times \frac{N}{2^i}$ matrix.
Thus,
{\small
\begin{align}
\Span\left({\bf H}_{N}\right) & = \Span\left(\left[{\bf F}_1 \;\;\;{\bf X}_{1}{\bf F}_1\right]\right) = \Span\left(\left[{\bf F}_2 \;\;\;{\bf X}_{2}{\bf F}_2 \;\;\;{\bf X}_{1}{\bf F}_2 \;\;\;{\bf X}_{1}{\bf X}_{2}{\bf F}_2\right]\right)\nonumber\\
&\hspace{0.17cm}\vdots\nonumber\\
&=\Span\left(\left\{\prod_{i = 1}^{\log_2(N)}{\bf X}_{i}^{x_i}{\bf w}: x_{i}\in \{0,1\}\right\}\right),
\end{align}
}
which proves the final part of Lemma 1.
\hfill$\Box$\\

\end{document}